\documentclass[reprint,amsmath,amssymb,aps,floatfix,superscriptaddress,showpacs,prl]{revtex4-1}

\usepackage{graphicx}% Include figure files
\usepackage{dcolumn}% Align table columns on decimal point
\usepackage{bm}% bold math% You should use BibTeX and apsrev.bst for references
\usepackage{color}
\usepackage{subfigure}
\usepackage{leftidx}
\usepackage{csquotes}
\MakeOuterQuote{"}
\usepackage{url}

\begin{document}
{

\title{Cosmic-ray Boron  Flux Measured from 8.4 GeV$/n$ to 3.8 TeV$/n$ \\
with the Calorimetric Electron Telescope \\ on the International Space Station}
%%%%%%%%%%%%%%%%%
\author{O.~Adriani}
\affiliation{Department of Physics, University of Florence, Via Sansone, 1 - 50019, Sesto Fiorentino, Italy}
\affiliation{INFN Sezione di Florence, Via Sansone, 1 - 50019, Sesto Fiorentino, Italy}
\author{Y.~Akaike}
\email[Corresponding author: ]{yakaike@aoni.waseda.jp}
\affiliation{Waseda Research Institute for Science and Engineering, Waseda University, 17 Kikuicho,  Shinjuku, Tokyo 162-0044, Japan}
\affiliation{JEM Utilization Center, Human Spaceflight Technology Directorate, Japan Aerospace Exploration Agency, 2-1-1 Sengen, Tsukuba, Ibaraki 305-8505, Japan}
\author{K.~Asano}
\affiliation{Institute for Cosmic Ray Research, The University of Tokyo, 5-1-5 Kashiwa-no-Ha, Kashiwa, Chiba 277-8582, Japan}
\author{Y.~Asaoka}
\affiliation{Institute for Cosmic Ray Research, The University of Tokyo, 5-1-5 Kashiwa-no-Ha, Kashiwa, Chiba 277-8582, Japan}
\author{E.~Berti} 
\affiliation{Department of Physics, University of Florence, Via Sansone, 1 - 50019, Sesto Fiorentino, Italy}
\affiliation{INFN Sezione di Florence, Via Sansone, 1 - 50019, Sesto Fiorentino, Italy}
\author{G.~Bigongiari}
\affiliation{Department of Physical Sciences, Earth and Environment, University of Siena, via Roma 56, 53100 Siena, Italy}
\affiliation{INFN Sezione di Pisa, Polo Fibonacci, Largo B. Pontecorvo, 3 - 56127 Pisa, Italy}
\author{W.R.~Binns}
\affiliation{Department of Physics and McDonnell Center for the Space Sciences, Washington University, One Brookings Drive, St. Louis, Missouri 63130-4899, USA}
\author{M.~Bongi}
\affiliation{Department of Physics, University of Florence, Via Sansone, 1 - 50019, Sesto Fiorentino, Italy}
\affiliation{INFN Sezione di Florence, Via Sansone, 1 - 50019, Sesto Fiorentino, Italy}
\author{P.~Brogi}
\affiliation{Department of Physical Sciences, Earth and Environment, University of Siena, via Roma 56, 53100 Siena, Italy}
\affiliation{INFN Sezione di Pisa, Polo Fibonacci, Largo B. Pontecorvo, 3 - 56127 Pisa, Italy}
\author{A.~Bruno}
\affiliation{Heliospheric Physics Laboratory, NASA/GSFC, Greenbelt, Maryland 20771, USA}
\author{J.H.~Buckley}
\affiliation{Department of Physics and McDonnell Center for the Space Sciences, Washington University, One Brookings Drive, St. Louis, Missouri 63130-4899, USA}
\author{N.~Cannady}
\affiliation{Center for Space Sciences and Technology, University of Maryland, Baltimore County, 1000 Hilltop Circle, Baltimore, Maryland 21250, USA}
\affiliation{Astroparticle Physics Laboratory, NASA/GSFC, Greenbelt, Maryland 20771, USA}
\affiliation{Center for Research and Exploration in Space Sciences and Technology, NASA/GSFC, Greenbelt, Maryland 20771, USA}
\author{G.~Castellini}
\affiliation{Institute of Applied Physics (IFAC),  National Research Council (CNR), Via Madonna del Piano, 10, 50019, Sesto Fiorentino, Italy}
\author{C.~Checchia}
\affiliation{Department of Physical Sciences, Earth and Environment, University of Siena, via Roma 56, 53100 Siena, Italy}
\affiliation{INFN Sezione di Pisa, Polo Fibonacci, Largo B. Pontecorvo, 3 - 56127 Pisa, Italy}
\author{M.L.~Cherry}
\affiliation{Department of Physics and Astronomy, Louisiana State University, 202 Nicholson Hall, Baton Rouge, Louisiana 70803, USA}
\author{G.~Collazuol}
\affiliation{Department of Physics and Astronomy, University of Padova, Via Marzolo, 8, 35131 Padova, Italy}
\affiliation{INFN Sezione di Padova, Via Marzolo, 8, 35131 Padova, Italy} 
%%%  family name- de Nolfo
\author{G.A.~de~Nolfo}
\affiliation{Heliospheric Physics Laboratory, NASA/GSFC, Greenbelt, Maryland 20771, USA}
%%%
\author{K.~Ebisawa}
\affiliation{Institute of Space and Astronautical Science, Japan Aerospace Exploration Agency, 3-1-1 Yoshinodai, Chuo, Sagamihara, Kanagawa 252-5210, Japan}
\author{A.~W.~Ficklin}
\affiliation{Department of Physics and Astronomy, Louisiana State University, 202 Nicholson Hall, Baton Rouge, Louisiana 70803, USA}
\author{H.~Fuke}
\affiliation{Institute of Space and Astronautical Science, Japan Aerospace Exploration Agency, 3-1-1 Yoshinodai, Chuo, Sagamihara, Kanagawa 252-5210, Japan}
\author{S.~Gonzi}
\affiliation{Department of Physics, University of Florence, Via Sansone, 1 - 50019, Sesto Fiorentino, Italy}
\affiliation{INFN Sezione di Florence, Via Sansone, 1 - 50019, Sesto Fiorentino, Italy}
\author{T.G.~Guzik}
\affiliation{Department of Physics and Astronomy, Louisiana State University, 202 Nicholson Hall, Baton Rouge, Louisiana 70803, USA}
\author{T.~Hams}
\affiliation{Center for Space Sciences and Technology, University of Maryland, Baltimore County, 1000 Hilltop Circle, Baltimore, Maryland 21250, USA}
\author{K.~Hibino}
\affiliation{Kanagawa University, 3-27-1 Rokkakubashi, Kanagawa, Yokohama, Kanagawa 221-8686, Japan}
\author{M.~Ichimura}
\affiliation{Faculty of Science and Technology, Graduate School of Science and Technology, Hirosaki University, 3, Bunkyo, Hirosaki, Aomori 036-8561, Japan}
\author{K.~Ioka}
\affiliation{Yukawa Institute for Theoretical Physics, Kyoto University, Kitashirakawa Oiwake-cho, Sakyo-ku, Kyoto, 606-8502, Japan}
\author{W.~Ishizaki}
\affiliation{Institute for Cosmic Ray Research, The University of Tokyo, 5-1-5 Kashiwa-no-Ha, Kashiwa, Chiba 277-8582, Japan}
\author{M.H.~Israel}
\affiliation{Department of Physics and McDonnell Center for the Space Sciences, Washington University, One Brookings Drive, St. Louis, Missouri 63130-4899, USA}
\author{K.~Kasahara}
\affiliation{Department of Electronic Information Systems, Shibaura Institute of Technology, 307 Fukasaku, Minuma, Saitama 337-8570, Japan}
\author{J.~Kataoka}
\affiliation{School of Advanced Science and	Engineering, Waseda University, 3-4-1 Okubo, Shinjuku, Tokyo 169-8555, Japan}
\author{R.~Kataoka}
\affiliation{National Institute of Polar Research, 10-3, Midori-cho, Tachikawa, Tokyo 190-8518, Japan}
\author{Y.~Katayose}
\affiliation{Faculty of Engineering, Division of Intelligent Systems Engineering, Yokohama National University, 79-5 Tokiwadai, Hodogaya, Yokohama 240-8501, Japan}
\author{C.~Kato}
\affiliation{Faculty of Science, Shinshu University, 3-1-1 Asahi, Matsumoto, Nagano 390-8621, Japan}
\author{N.~Kawanaka}
\affiliation{Yukawa Institute for Theoretical Physics, Kyoto University, Kitashirakawa Oiwake-cho, Sakyo-ku, Kyoto, 606-8502, Japan}
\author{Y.~Kawakubo}
\affiliation{Department of Physics and Astronomy, Louisiana State University, 202 Nicholson Hall, Baton Rouge, Louisiana 70803, USA}
\author{K.~Kobayashi}
\affiliation{Waseda Research Institute for Science and Engineering, Waseda University, 17 Kikuicho,  Shinjuku, Tokyo 162-0044, Japan}
\affiliation{JEM Utilization Center, Human Spaceflight Technology Directorate, Japan Aerospace Exploration Agency, 2-1-1 Sengen, Tsukuba, Ibaraki 305-8505, Japan}
\author{K.~Kohri} 
\affiliation{Institute of Particle and Nuclear Studies, High Energy Accelerator Research Organization, 1-1 Oho, Tsukuba, Ibaraki, 305-0801, Japan} 
\author{H.S.~Krawczynski}
\affiliation{Department of Physics and McDonnell Center for the Space Sciences, Washington University, One Brookings Drive, St. Louis, Missouri 63130-4899, USA}
\author{J.F.~Krizmanic}
\affiliation{Astroparticle Physics Laboratory, NASA/GSFC, Greenbelt, Maryland 20771, USA}
\author{P.~Maestro}
\email[Corresponding author: ]{maestro@unisi.it}
\affiliation{Department of Physical Sciences, Earth and Environment, University of Siena, via Roma 56, 53100 Siena, Italy}
\affiliation{INFN Sezione di Pisa, Polo Fibonacci, Largo B. Pontecorvo, 3 - 56127 Pisa, Italy}
\author{P.S.~Marrocchesi}
\affiliation{Department of Physical Sciences, Earth and Environment, University of Siena, via Roma 56, 53100 Siena, Italy}
\affiliation{INFN Sezione di Pisa, Polo Fibonacci, Largo B. Pontecorvo, 3 - 56127 Pisa, Italy}
\author{A.M.~Messineo}
\affiliation{University of Pisa, Polo Fibonacci, Largo B. Pontecorvo, 3 - 56127 Pisa, Italy}
\affiliation{INFN Sezione di Pisa, Polo Fibonacci, Largo B. Pontecorvo, 3 - 56127 Pisa, Italy}
\author{J.W.~Mitchell}
\affiliation{Astroparticle Physics Laboratory, NASA/GSFC, Greenbelt, Maryland 20771, USA}
\author{S.~Miyake}
\affiliation{Department of Electrical and Electronic Systems Engineering, National Institute of Technology (KOSEN), Ibaraki College, 866 Nakane, Hitachinaka, Ibaraki 312-8508, Japan}
\author{A.A.~Moiseev}
\affiliation{Department of Astronomy, University of Maryland, College Park, Maryland 20742, USA}
\affiliation{Astroparticle Physics Laboratory, NASA/GSFC, Greenbelt, Maryland 20771, USA}
\affiliation{Center for Research and Exploration in Space Sciences and Technology, NASA/GSFC, Greenbelt, Maryland 20771, USA}
\author{M.~Mori}
\affiliation{Department of Physical Sciences, College of Science and Engineering, Ritsumeikan University, Shiga 525-8577, Japan}
\author{N.~Mori}
\affiliation{INFN Sezione di Florence, Via Sansone, 1 - 50019, Sesto Fiorentino, Italy}
\author{H.M.~Motz}
\affiliation{Faculty of Science and Engineering, Global Center for Science and Engineering, Waseda University, 3-4-1 Okubo, Shinjuku, Tokyo 169-8555, Japan}
\author{K.~Munakata}
\affiliation{Faculty of Science, Shinshu University, 3-1-1 Asahi, Matsumoto, Nagano 390-8621, Japan}
\author{S.~Nakahira}
\affiliation{Institute of Space and Astronautical Science, Japan Aerospace Exploration Agency, 3-1-1 Yoshinodai, Chuo, Sagamihara, Kanagawa 252-5210, Japan}
\author{J.~Nishimura}
\affiliation{Institute of Space and Astronautical Science, Japan Aerospace Exploration Agency, 3-1-1 Yoshinodai, Chuo, Sagamihara, Kanagawa 252-5210, Japan}
\author{S.~Okuno}
\affiliation{Kanagawa University, 3-27-1 Rokkakubashi, Kanagawa, Yokohama, Kanagawa 221-8686, Japan}
\author{J.F.~Ormes}
\affiliation{Department of Physics and Astronomy, University of Denver, Physics Building, Room 211, 2112 East Wesley Avenue, Denver, Colorado 80208-6900, USA}
\author{S.~Ozawa}
\affiliation{Quantum ICT Advanced Development Center, National Institute of Information and Communications Technology, 4-2-1 Nukui-Kitamachi, Koganei, Tokyo 184-8795, Japan}
\author{L.~Pacini}
\affiliation{Department of Physics, University of Florence, Via Sansone, 1 - 50019, Sesto Fiorentino, Italy}
\affiliation{Institute of Applied Physics (IFAC),  National Research Council (CNR), Via Madonna del Piano, 10, 50019, Sesto Fiorentino, Italy}
\affiliation{INFN Sezione di Florence, Via Sansone, 1 - 50019, Sesto Fiorentino, Italy}
\author{P.~Papini}
\affiliation{INFN Sezione di Florence, Via Sansone, 1 - 50019, Sesto Fiorentino, Italy}
\author{B.F.~Rauch}
\affiliation{Department of Physics and McDonnell Center for the Space Sciences, Washington University, One Brookings Drive, St. Louis, Missouri 63130-4899, USA}
\author{S.B.~Ricciarini}
\affiliation{Institute of Applied Physics (IFAC),  National Research Council (CNR), Via Madonna del Piano, 10, 50019, Sesto Fiorentino, Italy}
\affiliation{INFN Sezione di Florence, Via Sansone, 1 - 50019, Sesto Fiorentino, Italy}
\author{K.~Sakai}
\affiliation{Center for Space Sciences and Technology, University of Maryland, Baltimore County, 1000 Hilltop Circle, Baltimore, Maryland 21250, USA}
\affiliation{Astroparticle Physics Laboratory, NASA/GSFC, Greenbelt, Maryland 20771, USA}
\affiliation{Center for Research and Exploration in Space Sciences and Technology, NASA/GSFC, Greenbelt, Maryland 20771, USA}
\author{T.~Sakamoto}
\affiliation{College of Science and Engineering, Department of Physics and Mathematics, Aoyama Gakuin University,  5-10-1 Fuchinobe, Chuo, Sagamihara, Kanagawa 252-5258, Japan}
\author{M.~Sasaki}
\affiliation{Department of Astronomy, University of Maryland, College Park, Maryland 20742, USA}
\affiliation{Astroparticle Physics Laboratory, NASA/GSFC, Greenbelt, Maryland 20771, USA}
\affiliation{Center for Research and Exploration in Space Sciences and Technology, NASA/GSFC, Greenbelt, Maryland 20771, USA}
\author{Y.~Shimizu}
\affiliation{Kanagawa University, 3-27-1 Rokkakubashi, Kanagawa, Yokohama, Kanagawa 221-8686, Japan}
\author{A.~Shiomi}
\affiliation{College of Industrial Technology, Nihon University, 1-2-1 Izumi, Narashino, Chiba 275-8575, Japan}
\author{P.~Spillantini}
\affiliation{Department of Physics, University of Florence, Via Sansone, 1 - 50019, Sesto Fiorentino, Italy}
\author{F.~Stolzi}
\affiliation{Department of Physical Sciences, Earth and Environment, University of Siena, via Roma 56, 53100 Siena, Italy}
\affiliation{INFN Sezione di Pisa, Polo Fibonacci, Largo B. Pontecorvo, 3 - 56127 Pisa, Italy}
\author{S.~Sugita}
\affiliation{College of Science and Engineering, Department of Physics and Mathematics, Aoyama Gakuin University,  5-10-1 Fuchinobe, Chuo, Sagamihara, Kanagawa 252-5258, Japan}
\author{A.~Sulaj} 
\affiliation{Department of Physical Sciences, Earth and Environment, University of Siena, via Roma 56, 53100 Siena, Italy}
\affiliation{INFN Sezione di Pisa, Polo Fibonacci, Largo B. Pontecorvo, 3 - 56127 Pisa, Italy}
\author{M.~Takita}
\affiliation{Institute for Cosmic Ray Research, The University of Tokyo, 5-1-5 Kashiwa-no-Ha, Kashiwa, Chiba 277-8582, Japan}
\author{T.~Tamura}
\affiliation{Kanagawa University, 3-27-1 Rokkakubashi, Kanagawa, Yokohama, Kanagawa 221-8686, Japan}
\author{T.~Terasawa}
\affiliation{Institute for Cosmic Ray Research, The University of Tokyo, 5-1-5 Kashiwa-no-Ha, Kashiwa, Chiba 277-8582, Japan}
\author{S.~Torii}
%\email[]{torii.shoji@waseda.jp}
\affiliation{Waseda Research Institute for Science and Engineering, Waseda University, 17 Kikuicho,  Shinjuku, Tokyo 162-0044, Japan}
\author{Y.~Tsunesada}
\affiliation{Graduate School of Science, Osaka Metropolitan University, Sugimoto, Sumiyoshi, Osaka 558-8585, Japan }
\affiliation{ Nambu Yoichiro Institute for Theoretical and Experimental Physics, Osaka Metropolitan University,  Sugimoto, Sumiyoshi, Osaka  558-8585, Japan}
\author{Y.~Uchihori}
\affiliation{National Institutes for Quantum and Radiation Science and Technology, 4-9-1 Anagawa, Inage, Chiba 263-8555, Japan}
\author{E.~Vannuccini}
\affiliation{INFN Sezione di Florence, Via Sansone, 1 - 50019, Sesto Fiorentino, Italy}
\author{J.P.~Wefel}
\affiliation{Department of Physics and Astronomy, Louisiana State University, 202 Nicholson Hall, Baton Rouge, Louisiana 70803, USA}
\author{K.~Yamaoka}
\affiliation{Nagoya University, Furo, Chikusa, Nagoya 464-8601, Japan}
\author{S.~Yanagita}
\affiliation{College of Science, Ibaraki University, 2-1-1 Bunkyo, Mito, Ibaraki 310-8512, Japan}
\author{A.~Yoshida}
\affiliation{College of Science and Engineering, Department of Physics and Mathematics, Aoyama Gakuin University,  5-10-1 Fuchinobe, Chuo, Sagamihara, Kanagawa 252-5258, Japan}
\author{K.~Yoshida}
\affiliation{Department of Electronic Information Systems, Shibaura Institute of Technology, 307 Fukasaku, Minuma, Saitama 337-8570, Japan}
\author{W.~V.~Zober}
\affiliation{Department of Physics and McDonnell Center for the Space Sciences, Washington University, One Brookings Drive, St. Louis, Missouri 63130-4899, USA}

\collaboration{CALET Collaboration}

\date{\today}

\begin{abstract}
We present the measurement of the energy dependence of the boron flux in cosmic rays and  its ratio to the carbon flux  \textcolor{black}{in an energy interval   from 8.4 GeV$/n$ to 3.8 TeV$/n$}
based on the data collected by the CALorimetric Electron Telescope (CALET) during $\sim 6.4$ years of operation on the International Space Station.
An update of the energy spectrum of carbon is also presented with an increase in statistics over our previous measurement.  
The observed boron flux shows a spectral hardening at the same transition energy $E_0 \sim 200$ GeV$/n$ of the C spectrum, though B and C fluxes have  different energy dependences. 
The spectral index of the B spectrum  is found to be $\gamma = -3.047\pm0.024$ in the interval $25 < E <  200$ GeV$/n$.
The B spectrum hardens by $\Delta \gamma_B=0.25\pm0.12$, while the best fit value for the spectral variation of C  is $\Delta \gamma_C=0.19\pm0.03$.
The B/C flux ratio is compatible with a hardening of $0.09\pm0.05$, 
though a single power-law energy dependence  cannot be ruled out  given the current statistical uncertainties.
A break in the B/C ratio energy dependence would support the recent AMS-02 observations that secondary cosmic rays exhibit a stronger hardening  than primary ones.
We also perform a fit to the B/C ratio with  a leaky-box model of the cosmic-ray propagation in the Galaxy in order to probe a possible residual value $\lambda_0$ of the 
mean escape path length $\lambda$ at high energy. 
We find that our B/C data are compatible with a non-zero value of  $\lambda_0$, which can be 
interpreted as the column density of matter that cosmic rays cross within the acceleration region. 
\end{abstract}

\maketitle

\section{Introduction}
The larger relative abundance of light elements such as Li, Be, B  
 in cosmic rays (CR) compared to the solar system abundance
is a proof of their secondary origin. They are produced by the spallation reactions 
of primary CR, injected and accelerated in astrophysical sources, with nuclei of the interstellar medium (ISM).
Measurements of the secondary-to-primary abundance ratios (as B/C) 
make it possible to probe galactic propagation models and  constrain their parameters, since they are expected 
 to be proportional at high energy to  the average amount of material $\lambda$ 
 traversed by CR in the Galaxy, which in turn is inversely proportional to the CR diffusion coefficient $D$.
 Earlier measurements \cite{HEAO, CRN-BC, CREAM1-BC, TRACER2012, PAMELA-C}
 indicate that $\lambda$ decreases with increasing CR energy per nucleon $E$, 
following a power-law $\lambda \propto E^{-\delta}$, where $\delta$ is the diffusion spectral index. 
The recently observed hardening in the spectrum of CR of different nuclear species 
\cite{AMS-BC, AMS-LiBeB, AMS-PhysRep, CALET-CO, CALET-P2, DAMPEp, DAMPEhe} 
can be explained as due to subtle effects of CR transport including: 
an inhomogeneous or an energy-dependent diffusion coefficient \cite{Tomassetti, Aloisio2015,Johannesson};
the possible re-acceleration of secondary particles when they occasionally cross a supernova shock during propagation \cite{Cuoco}; 
and/or the production of a small fraction of secondaries by interactions of primary nuclei with matter (source grammage)
inside the acceleration region \cite{Cowsik, Bresci, Evoli}. 
To investigate these phenomena, a precise determination of the energy dependence of $\lambda$
is needed. 
That can be achieved  by extending the measurements of secondary CR
in the TeV$/n$ region with high statistics and reduced systematic uncertainties.
In this Letter, we present new direct measurements of the energy spectra of  boron, carbon and  of the boron-to-carbon ratio
in the energy range  from 8.4 GeV$/n$ to 3.8 TeV$/n$,
based on the data collected by the CALorimetric Electron Telescope (CALET)
 \cite{CALET, CALET-ELE2018, CALET-PROTON} from October 13, 2015 to February 28, 2022 aboard the International Space Station (ISS).
\section{Detector}
The CALET instrument comprises a CHarge Detector (CHD),
a finely segmented pre-shower IMaging Calorimeter (IMC), and a Total AbSorption Calorimeter (TASC).
A complete description of the instrument can be found in the Supplemental Material (SM) of Ref.~\cite{CALET-ELE2017}.

The IMC consists of 7 tungsten plates interspaced with eight double layers of 
scintillating fibers, arranged  
along orthogonal directions.
Fiber signals are used to reconstruct
the CR particle trajectory 
by applying 
a combinatorial Kalman filter \cite{paolo2017}. 
The estimated error in the determination of the arrival direction of B and C nuclei is $\sim0.1^\circ$ 
with a corresponding spatial resolution of the impact point on the CHD of $\sim$220 $\mu$m.

The identification of the particle charge $Z$ is based  on the measurements of the ionization deposits in the CHD and IMC.
The CHD, located above the IMC, is comprised of two hodoscopes (CHDX, CHDY) made of 14 plastic scintillator paddles each, 
arranged perpendicularly to each other. 
The particle trajectory is used to identify the CHD paddles and IMC fibers traversed by the primary particle
and to determine the path length correction to be applied to the signals to extract samples of the ionization energy loss ($dE/dx$). 
Three charge values ($Z_{\rm{CHDX}}$, $Z_{\rm{CHDY}}$, $Z_{\rm{IMC}}$) are reconstructed, on an event-by-event basis,  from the measured $dE/dx$ in each CHD layer 
and the average of the $dE/dx$ samples  along the track in the top half of IMC \cite{CALET-CO}. 
The CHD can resolve individual chemical elements from $Z=1$ to 40, %\textcolor{red}{with excellent charge resolution,}
while the saturation of the fiber signals limits the IMC charge measurement %range 
to $Z \lesssim 14$.
 \textcolor{black}{The charge resolution of the CHD (IMC) is $\sim0.15\, (0.24)\, e$ (charge unit) 
in the elemental range from B to O.}

The TASC is a homogeneous calorimeter made of 12 layers of %16 
lead-tungstate  bars, each 
read out by photosensors and a front-end electronics spanning a dynamic range $>10^6$. 
The total thickness of the instrument is equivalent to 30 radiation lengths and 1.3 proton nuclear interaction lengths. 

The TASC was calibrated %response was studied 
at the CERN SPS in 2015 using a beam of accelerated ion fragments with $A/Z = 2$ and kinetic energy of 13, 19 and 150 GeV$/n$ \cite{akaike2015}.
The response curve for interacting particles of each nuclear species
is nearly gaussian at a fixed beam energy.
The mean energy released in the TASC is $\sim$20\% of the particle energy and the resolution is close to 30\%. 
The energy response of the TASC turned out to be linear up to the maximum particle energy (6 TeV) available at the beam, 
as described in the SM of Ref.~\cite{CALET-CO}.

Monte Carlo (MC) simulations,  reproducing the detailed detector configuration, physics processes, as well as detector signals,
are based on the EPICS simulation package~\cite{EPICS} and employ 
the hadronic interaction model DPMJET-III~\cite{dpmjet3prl}. Independent simulations based on 
Geant4 10.5~\cite{G4} are used to assess the systematic uncertainties.
\section{Data analysis}
We have analyzed flight data (FD) collected in 2331 days of CALET operation aboard the ISS. 
Raw data are corrected for non-uniformity in light output, time and temperature dependence, and gain differences among the channels
by using penetrating protons and He particles selected by a dedicated trigger mode \cite{CALET2018}.
Correction curves for the reduction of the scintillator light yield due to the quenching effect in the CHD and IMC
are obtained from FD  by fitting subsets for each nuclear species to a function of $Z^2$ using a \enquote{halo} model \cite{GSI}.

Boron and carbon candidates are searched for among events selected by the onboard high-energy (HE) shower trigger,  
which requires the coincidence of the summed signals of the last two IMC double layers  and the top TASC layer. 
The total observation live time for  the HE trigger is 
\textcolor{black}{$T=4.72\times10^4$} hours, corresponding to 87.2\% of the total observation time. 
In order to mitigate the effect of possible temporal variations of the  trigger thresholds on the trigger efficiency, 
an offline trigger is applied to FD with higher thresholds than the onboard trigger.
Triggered  particles entering the instrument from lateral sides 
or late-interacting  in the lower half of the calorimeter 
are rejected based on the large fraction of energy leakage estimated from the shape of the longitudinal and lateral shower profiles.
All reconstructed events with one well-fitted track 
passing through the top surface of the CHD and the bottom surface of the TASC (excluding a border region of 2 cm) are then selected. 
The geometrical acceptance  for this category of events is $S\Omega \sim$510 cm$^2$sr.  

Boron and carbon candidates are identified 
by applying  window charge cuts of half-width $0.45\, e$ centered on the nominal  values ($Z=5, 6$) to the 
distribution of the average charge in the CHD ($Z_{\rm{CHD}}$)
obtained after requiring that 
$Z_{\rm{CHDX}}$ and $Z_{\rm{CHDY}}$ are consistent with each other within 10\% and 
$|Z_{\rm{CHD}} - Z_{\rm{IMC}}|< 1$, as shown in Fig. S2 of the SM \cite{PRL-SM}. 
The consistency of the charge values measured by each of the four upper IMC fiber layers is also required.

An additional cut on the track width ($\rm{TW}$) is applied to reject particles undergoing a charge-changing nuclear interaction in the upper part of the instrument.
The $\rm{TW}$ variable is defined as  the difference, normalized to the particle charge, 
 between the total energy deposited in the clusters of nearby fibers crossed by the reconstructed track 
and the sum of the fiber signals in the cluster cores. 
Examples of $\rm{TW}$ distributions are shown in Fig. S3 of the SM \cite{PRL-SM}.

The field-of-view (FOV) of CALET at large zenith angle ($>$45$^\circ$) is partially shielded by fixed structures on the ISS. 
Moreover, moving structures (e.g. solar panels, robotic arms) can cross
the FOV for short periods of time during ISS operations.
CR interactions in these structures can create secondary nuclei that, if detected by CALET, may induce 
a contamination in the flux measurements. 
To avoid that, the events \textcolor{black}{($\sim$8\% of the final candidate samples)} with reconstructed trajectories pointing to obstacles in the FOV are discarded in the analysis. 

With this selection procedure 
1.99$\times 10^5$ B and  9.27$\times 10^5$ C nuclei are identified. 
For flux measurements, 
an iterative unfolding Bayesian method \cite{Ago} 
is applied to correct the distributions (Fig. S4 of the SM \cite{PRL-SM}) 
of the total energy deposited in the TASC ($E_{\rm TASC}$) 
for significant bin-to-bin migration effects 
(due to the limited energy resolution) and infer the primary particle energy. 
The response matrix for the unfolding procedure 
is derived using MC simulations after applying the same selection procedure as for FD.
The energy spectrum is obtained from the unfolded energy distribution as follows:
\begin{equation}
\Phi(E) = \frac{N(E)}{\Delta E\;  \varepsilon(E) \;  S\Omega \;  T }
\label{eq_flux}
\end{equation}
\begin{equation}
N(E) = U \left[N_{obs}(E_{\rm TASC}) - N_{bg}(E_{\rm TASC}) \right]
\end{equation}
where: $\Delta E$ is the energy bin width;
$E$ the kinetic energy per nucleon calculated as the geometric mean of the lower and upper bounds of the bin; 
$N(E)$ the bin content in the unfolded distribution;
$\varepsilon (E)$ the total selection efficiency (Fig.~S5 of the SM \cite{PRL-SM}); 
$U()$  the iterative unfolding procedure; 
$N_{obs}(E_{\rm TASC})$ the bin content of the observed energy distribution (including background);
$N_{bg}(E_{\rm TASC})$ the bin content of background events in the observed energy distribution. 
The background contamination in the final B sample is estimated from $\rm{TW}$ distributions in different intervals of  $E_{\rm TASC}$, after applying the complete charge selection procedure.
The contamination fraction $N_{bg}/N_{obs}$ is $\sim 1\%$ for $E_{\rm TASC}<10^2$ GeV and grows logarithmically with $E_{\rm TASC}$ for $E_{\rm TASC}>10^2$ GeV, 
approaching $\sim 7\%$ at  1.5 TeV. The background is negligible for C. \\\\
%%%%%%%%%%%%%%%%%%%%%%%%%%%
%%%%%%%%%%%%%%%%%%%%%%%%%%%
\section{Systematic Uncertainties}
\vspace{-0.2cm}
Different sources of systematic uncertainties were studied, including
trigger efficiency, charge identification, energy scale, unfolding procedure, MC simulations, B isotopic composition, and background subtraction.

The HE trigger efficiency  was  measured  as a function of $E_{\rm TASC}$ using a subset of data taken with a minimum bias trigger.  
\textcolor{black}{The small differences ($<$1\%) found between the HE efficiency curves and the predictions from MC simulations  (Fig.~S1 of the SM \cite{PRL-SM})
induce a systematic error of $\pm 0.8\%$  ($\pm 0.7\%$) in the B (C) flux. }

The systematic error related to charge identification was studied by varying 
the width of  the window cuts for $Z_{\rm{CHD}}$ between 0.43\,$e$ and 0.47\,$e$  and 
the boundary $\alpha$ of the consistency  cut 
$|Z_{\rm{CHD}} - Z_{\rm{IMC}}|< \alpha$
between 0.9 and 1.1. 
The result was a flux variation ranging from $-1.1\%$ to $3.1\%$ for B, 
and $-1.5\%$ to $0.9\%$ for C,  depending on the energy bin. 

\textcolor{black}{The uncertainty ($\pm2$\%) in the energy scale  from the beam test calibration 
affects the absolute normalization of the B and C spectra by $\pm$3\% but not their shape.}

The uncertainty due to the unfolding procedure 
was evaluated by using  different response matrices computed by varying the spectral index 
of the generation spectrum of MC simulations.
The resulting error in the absolute flux is $\pm 1.5\%$ for B and $\pm 0.5\%$ for C.

Since it is not possible to validate MC simulations with  beam test data in the high-energy region,
a comparison between different MC programs,  i.e. EPICS and Geant4,  was performed. 
We found that the  selection efficiencies are similar, 
but the energy response matrices differ significantly in the low and high energy regions.
The resulting fluxes for B (C) show discrepancies not exceeding 6\% (10\%) below 20 GeV$/n$
and 12\% (10\%) above 300 GeV$/n$, respectively.
This is the dominant source of systematic uncertainties.

The uncertainty of the residual background contamination leads to a maximum error of $3\%$ in the B flux above 400 GeV$/n$, and $\le 2\%$ below. 

Since CALET cannot distinguish among the B isotopes, 
the spectral binning in 
kinetic energy per nucleon is calculated assuming
an isotopic composition of 70\% of $^{11}$B
and 30\% of $^{10}$B as in Ref.~\cite{AMS-BC}.
We checked with MC that a variation of $\pm$10\% in the abundance of  $^{11}$B
causes a $\pm$1\% difference in the selection efficiency and a $\mp$1.7\% change in the flux normalization.

Other energy-independent systematic uncertainties affecting the normalization
include live time (3.4\%, as explained in the SM of Ref.~\cite{CALET-ELE2017}) and long-term stability of charge calibration ($0.5\%$). 

The energy dependence of all the systematic uncertainties  is shown in  Fig.~S6 of the SM~\cite{PRL-SM}. 
\textcolor{black}{Finally, an independent analysis, using different tracking and charge identification procedures \cite{akaike2021}, turned out  to be in very good agreement
with the results reported in this Letter. }
%%%%%%%%%%%%%%%%%%%%%%%%%%%%%%%%%%%%%%%%
\section{Results}
The energy spectra of B and C and their flux ratio measured with CALET 
are shown in Fig.~\ref{fig:flux};
the corresponding data tables including statistical and systematic errors are reported in the SM \cite{PRL-SM}.
CALET spectra are compared with results from space-based  \cite{HEAO,CRN-BC,PAMELA-C, AMS-PhysRep, AMS-LiBeB} 
and balloon-borne \cite{CREAM1-BC, TRACER2012, ATIC, CREAM2} experiments.
%%%%%%%%%%%%%%%%%%%%%%%%%%%%%%%%%%%%%%%%%%%%
\begin{figure} \centering
\includegraphics[width=\hsize]{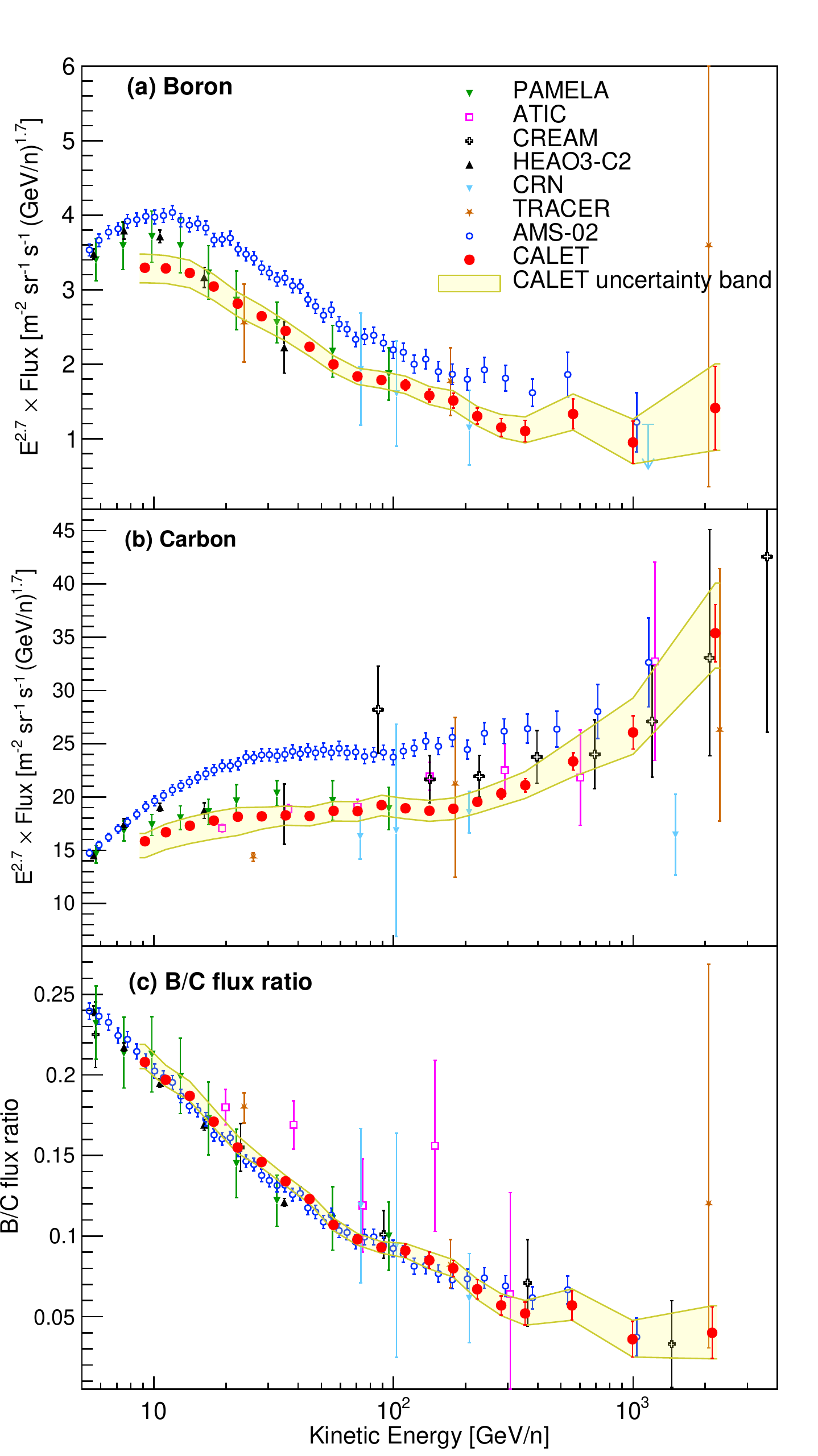} 
\caption{\scriptsize CALET (a) boron and (b) carbon flux (multiplied by $E^{2.7}$) and (c)  ratio of boron to carbon, as a function of kinetic energy per nucleon $E$.
Error bars of CALET data (red) represent the statistical uncertainty only, while the yellow band indicates the quadratic sum of statistical and systematic errors. 
Also plotted are other direct measurements \cite{HEAO, CRN-BC, CREAM1-BC, TRACER2012, PAMELA-C, AMS-PhysRep, AMS-LiBeB, ATIC, CREAM2}.
An enlarged version of the figure is available in Fig.~S8 of the SM \cite{PRL-SM}. }
\label{fig:flux}
\end{figure}\noindent
%%%%%%%%%%%%%%%%%%%%%%%%%%%%%%%%%%%%%%%%%%%%%%%%%%%
The B spectrum is  consistent with that of PAMELA \cite{PAMELA-C} and most  of the earlier experiments 
but the absolute normalization is in tension with that of AMS-02, 
as already pointed out by our previous measurements of the C, O and Fe fluxes  \cite{CALET-CO, CALET-Fe}.
However we notice that the B/C ratio (Fig.~\ref{fig:flux}(c)) is consistent with the one measured by AMS-02. 
The C spectrum shown here is based on a larger dataset but it is consistent with our earlier result
and includes an improved assessment of systematic errors.

Figure \ref{fig:BCfit} shows the fits to CALET B and C data with a double power-law function (DPL)
\begin{equation}
\Phi(E) = \begin{cases} c \left(\frac{E}{\text{GeV}} \right)^{\gamma} & E\le E_0\\
c \left(\frac{E}{\text{GeV}} \right)^{\gamma}  \left(\frac{E}{E_0}\right)^{\Delta\gamma}   & E>E_0 \end{cases}
\label{eq:DPL}
\end{equation}
where $c$ is a normalization factor, $\gamma$ the spectral index, and 
$\Delta \gamma$ the spectral index change above the transition energy $E_0$. 
A single power-law function (SPL) is also shown for comparison, where $\Delta\gamma = 0$ is fixed in Eq.~(\ref{eq:DPL})  %SPL, Eq.~2 in SM \cite{PRL-SM}))  fitted to data 
and the fit is limited to data points with $25 < E <  200$ GeV$/n$ and extrapolated above.
The DPL fit to the C spectrum  in the energy range [25, 3800] GeV$/n$
yields $\gamma_C = -2.670\pm0.005$ and a spectral index increase $\Delta\gamma_C=0.19\pm0.03$ 
at $E_0^C = (220\pm20)$ GeV$/n$ confirming our first results reported in Ref.~\cite{CALET-CO}.
For the B spectrum,  the parameter $E_0^B$ is fixed to the fitted value of $E_0^C$. 
The best fit parameters for B are:  $\gamma_B = -3.047\pm0.024$ and  $\Delta\gamma_B=0.25\pm0.12$  %$E_0^B = (240\pm40)$ GeV$/n$ 
with  $\chi^2/$d.o.f. = 11.9/12.
\textcolor{black}{The energy spectra are clearly different as expected for primary and secondary CR, 
and the fit results seem to indicate, albeit with low statistical significance, that the flux hardens more for B than for C above 200 GeV$/n$.
A similar indication also comes from the fit to the B/C flux ratio (Fig.~\ref{fig:BCratio}). 
In the energy range [25, 3800] GeV$/n$,
it can be fitted
with a SPL function with  spectral index $\Gamma = -0.366\pm0.018$ ($\chi^2/$d.o.f. = 9.4/13).
However a DPL function provides a better fit suggesting a trend of the data towards a flattening of the B/C ratio at high energy, 
with a  spectral index change $\Delta\Gamma = 0.09\pm0.05$ ($\chi^2/$d.o.f. = 8.7/12) 
above $E_0^C$, which is left as a fixed parameter in the fit. 
This result is  consistent with that of AMS-02 \cite{AMS-LiBeB}, and supports 
the hypothesis  that secondary B exhibits a stronger hardening  than primary C, 
although no definitive conclusion can be drawn due to the large uncertainty in $\Delta\Gamma$  given by our present statistics.}
\begin{figure} \centering
\includegraphics[scale=0.46]{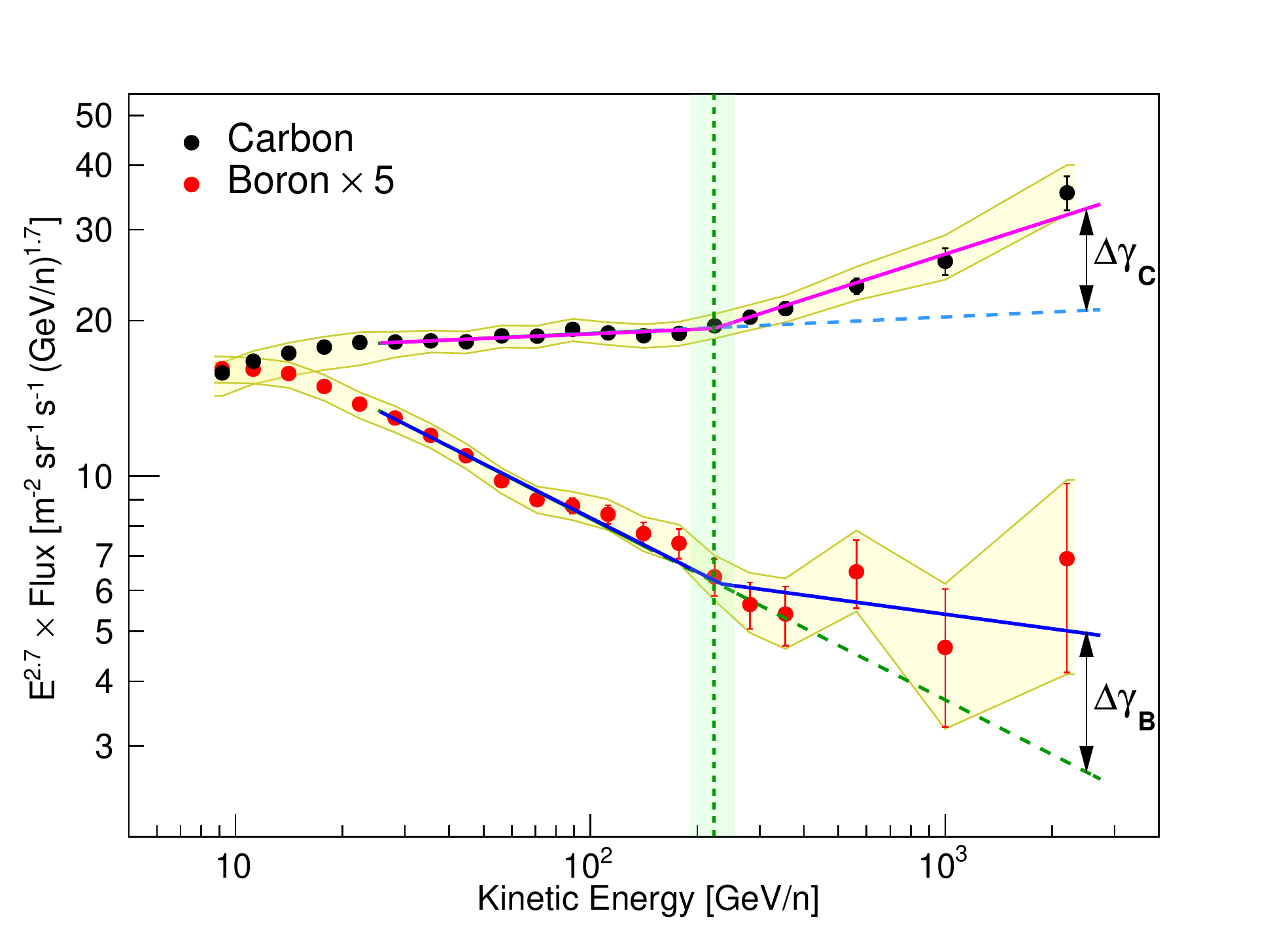}
\caption{\scriptsize  CALET  B (red dots)  and C (black dots) energy spectra are fitted with DPL  functions (magenta line for C,  blue line for B) in the energy range [25, 3800] GeV$/n$. 
The B spectrum is multiplied by a factor 5 to overlap the low-energy region of the C spectrum.
The dashed  lines represent the extrapolation of a SPL function fitted to data in the energy range [25, 200] GeV$/n$.
$\Delta\gamma$ is the change of the spectral index above the transition energy $E_0^C$ (from the fit to C data), represented by the vertical green dashed line.
The green band shows the error interval of $E_0^C$.}
\label{fig:BCfit}
\end{figure}\noindent
%%%%%%%%%%%%%%%%%
%%%%%%%%%%%%%%%%%%%%%%%%%%%
\begin{figure} 
%\vspace{0.6cm}
\centering
\includegraphics[scale=0.46]{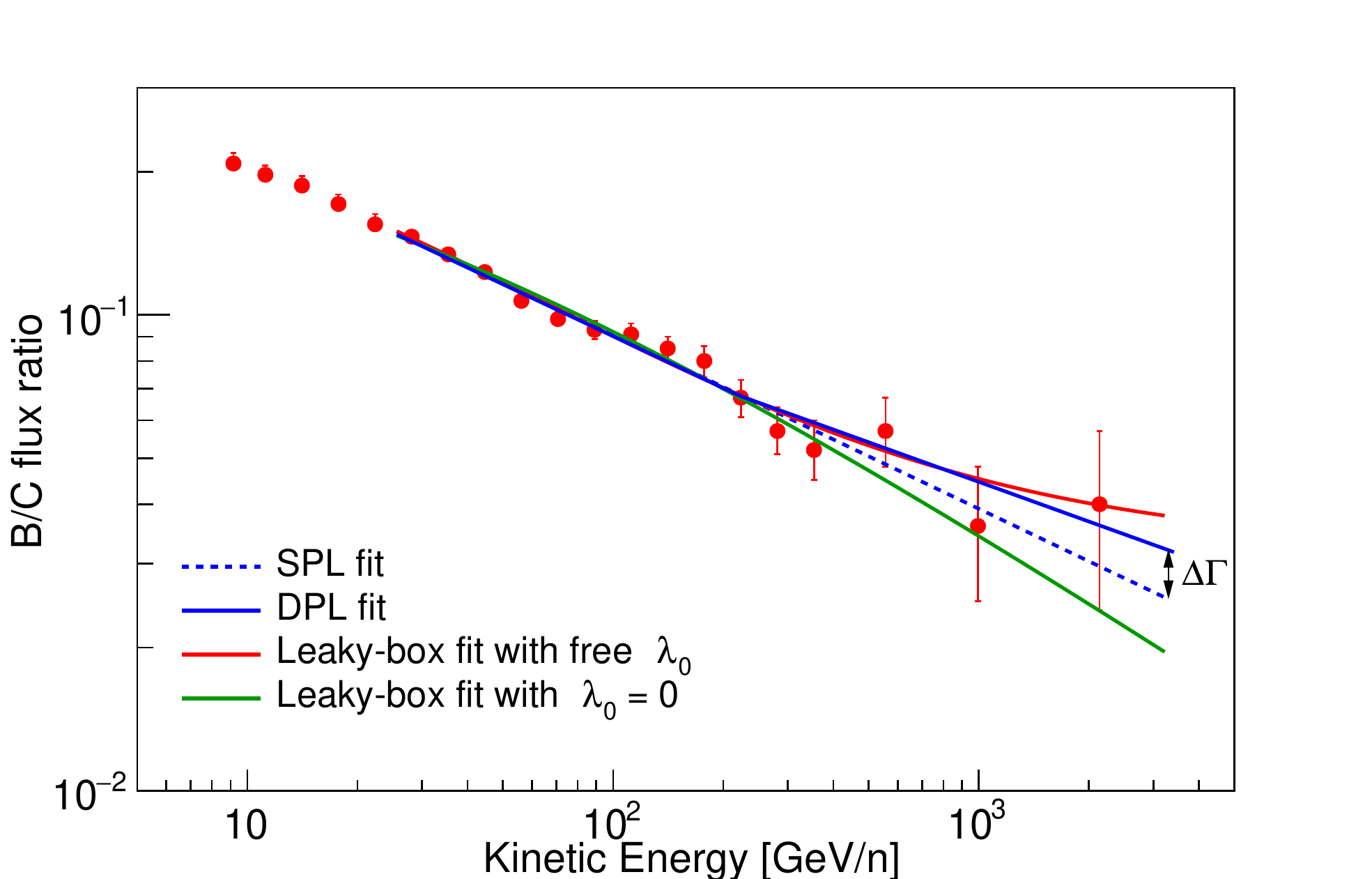}
\caption{\scriptsize  The CALET B/C ratio  fitted to  different functions. The error bars are the sum in quadrature of statistical and systematic uncertainties.
The data are fitted to a DPL (solid blue line) and a SPL (dashed blue line) function  in the energy interval  [25, 3800] GeV$/n$.
The red and green lines represent the fitted functions from a leaky-box model (Eq.~(\ref{eq:LB})) with the $\lambda_0$ parameter left free to vary and fixed to zero, respectively.}
\label{fig:BCratio}
\end{figure}\noindent
%%%%%%%%%%%%%%%%%

\textcolor{black}{Within the  "leaky-box" (LB) approximate modeling of the particle transport in the Galaxy \cite{TRACER2012}, the B/C flux ratio can be expressed as} 
%%%
\begin{equation}
\frac{\Phi_B(E)}{\Phi_C(E)}= \frac{\lambda(E) \lambda_B}{\lambda(E) + \lambda_B}   \left[ \frac{1}{\lambda_{C\rightarrow B}}  + \frac{\Phi_O(E)}{\Phi_C(E)} \frac{1}{\lambda_{O\rightarrow B}}\right]
\label{eq:LB}
\end{equation}
%%%%
where $\lambda_B$ is the interaction length of B nuclei with matter of the ISM and 
$\lambda_{C\rightarrow B}$ ($\lambda_{O\rightarrow B}$) is the average path length for a nucleus C (O) to spall into B.
The spallation path lengths are calculated using the parametrization of the total and partial charge changing cross sections provided in Ref.~\cite{Webber}, 
assuming that they are constant above a few GeV$/n$. 
The $\Phi_C(E)/\Phi_O(E)$ ratio is measured to be independent of energy and close to 0.91 \cite{CALET-CO}.
The contribution due to  the spallation of heavier primary nuclei (Ne, Mg, Si, Fe) to the B flux is estimated to be $\sim$10\% of 
the C+O flux and therefore it was not taken into account in Eq.~(\ref{eq:LB}).
Assuming a composition of the ISM of 90\% hydrogen and 10\% helium, 
we calculate $\lambda_B$=9.4 g/cm$^2$, while    the constant term enclosed in square brackets in Eq.~(\ref{eq:LB}) is 27 g/cm$^2$.

The LB model describes the diffusion of CR in the Galaxy with
a mean escape path length $\lambda(E)$ which, according to presently available direct measurements, is parametrized  as a power-law function of kinetic energy $E$ as follows:
\begin{equation}
\lambda(E) = k E^{-\delta} + \lambda_0
\label{eq:lambda}
\end{equation}
where $\delta$ is the diffusion coefficient spectral index.
A residual path length $\lambda_0$ is included in the  asymptotic behavior of $\lambda$. 
It can be interpreted as the amount of matter traversed by CR inside the acceleration region (source grammage).  
Fitting our B/C data to Eq.~(\ref{eq:LB}) (Fig.~\ref{fig:BCratio}), 
the  best fit values  without the source grammage term ($\lambda_0 = 0$) are:
$ k = 11.2 \pm 0.5$  g/cm$^2$,   $\delta= 0.52 \pm 0.02$  ($\chi^2/$d.o.f. = 13.6/13).
Leaving  instead $\lambda_0$ free to vary in the LB fit, we obtain:
$ k = 12.0 \pm 0.9$  g/cm$^2$,   $\delta= 0.71 \pm 0.11$, $\lambda_0 = 0.95 \pm 0.35$  g/cm$^2$ ($\chi^2/$d.o.f. = 9.6/12).
These results  suggest the possibility of a non-null value of the residual path length (though with a large uncertainty) which could be the cause of the apparent flattening of the B/C ratio  
at high energy. The best fit values of $\delta$ and $\lambda_0$ are compatible with the ones obtained from
 a combined analysis of the B/C data from earlier experiments \cite{TRACER2012},
and with the predictions of some recent theoretical works \cite{Cuoco,Evoli}.
%%%%%%%%%%%%%%%%%%%%%%%%%%%%%%%%%%%
\section{Conclusion}
The CR boron spectrum  has been measured by CALET up to 3.8 TeV$/n$ using 76.5 months of data collected aboard the ISS.
\textcolor{black}{Our observations show that, despite their different energy dependence, boron  and carbon fluxes exhibit a spectral hardening
occurring at about the same energy. 
Within the limitations of our data's present statistical significance, the boron  spectral index change is found to be slightly larger than that of  carbon.
This trend seems to corroborate 
the hypothesis that secondary CR harden more than the primaries, as recently reported by AMS-02 \cite{AMS-LiBeB}.}
Interpreting our data with a LB model, we argue 
that the trend of the energy dependence of the B/C ratio in the TeV$/n$ region 
could suggest  a possible presence of a residual propagation path length, compatible with 
the hypothesis that a fraction of secondary B nuclei can be produced near the CR source. 
\section{Acknowledgments}
\begin{acknowledgments}
We gratefully acknowledge JAXA's contributions to the development of CALET and to the operations onboard the International Space Station. 
We also express our sincere gratitude to ASI and NASA for their support of the CALET project. 
This work was supported in part by JSPS Grant-in- Aid for Scientific Research (S) Grant No. 19H05608, 
 JSPS Grand-in-Aid for Scientific Research (C) No. 21K03592
and by the MEXT-Supported Program for the Strategic 
Research Foundation at Private Universities (2011–2015) (Grant No. S1101021) at Waseda University. 
The CALET effort in Italy is supported by ASI under Agreement No. 2013-018-R.0 and its amendments. 
The CALET effort in the U.S. is supported by NASA through Grants No. 80NSSC20K0397, No. 80NSSC20K0399, and No. NNH18ZDA001N-APRA18-004.
\end{acknowledgments}
%

%%%%%%%%%% Merge with supplemental materials %%%%%%%%%%
%\pagebreak
%%%%%%%%%%%%%%%%%%%%%%%%%%%%%%%%%%%%%%%%%%%%%%%%%%%%%%%
\widetext
\clearpage
\begin{center}
%\textbf{\large Supplemental Materials: Title for main text}
\end{center}
%%%%%%%%%% Merge with supplemental materials %%%%%%%%%%
%%%%%%%%%% Prefix a "S" to all equations, figures, tables and reset the counter %%%%%%%%%%
\setcounter{equation}{0}
\setcounter{figure}{0}
\setcounter{table}{0}
\setcounter{page}{1}
\makeatletter
\renewcommand{\theequation}{S\arabic{equation}}
\renewcommand{\thefigure}{S\arabic{figure}}
\renewcommand{\bibnumfmt}[1]{[S#1]}
\renewcommand{\citenumfont}[1]{S#1}
%%%%%%%%%% Prefix a "S" to all equations, figures, tables and reset the counter %%%%%%%%%%
%%%%%%%%%%%%%%%%%%%%%%%%%%%%%%%%%%%%%%%%%%%%%%%%%%%%%%%
\begin{center}
\textbf{\large The Cosmic-ray Boron  Flux Measured from 8.4 GeV/n to 3.8 TeV/n \\
with the Calorimetric Electron Telescope \\ on the International Space Station\\
\vspace*{0.2cm}
SUPPLEMENTAL MATERIAL}	\\
\vspace*{0.2cm}
(CALET collaboration) 
\end{center}
\vspace*{1cm}
Supplemental material concerning \enquote{The Cosmic-ray Boron  Flux Measured from 8.4 GeV/n to 3.8 TeV/n 
with the Calorimetric Electron Telescope  on the International Space Station}.
\vspace*{1cm}
%%%%%%%%%%%%%%%%%
\clearpage
\section{ADDITIONAL INFORMATION ON THE DATA ANALYSIS}
{\bf Trigger.}  The high-energy (HE) trigger efficiency was measured directly from the flight data (FD) by using dedicated runs where in addition to HE, 
a low-energy (LE) trigger was active. 
The  trigger logic is the same for both triggers (i.e. coincidence of the pulse heights  of the  last two pairs of IMC layers and the top TASC layer) 
but lower discriminator thresholds are set for the input signals in the case of the LE trigger, 
allowing the instrument to trigger on  penetrating nuclei with $Z>2$.
The ratio of the number of events counted by both  triggers to those recorded  by the LE trigger alone
is  an estimate  of the HE trigger efficiency in each bin of deposited energy.
The HE trigger efficiency curves as a function of the total deposited energy in the TASC ($E_{\rm{TASC}}$)  are shown in Fig. \ref{fig:HEeff},
where they are compared with MC simulations in which both  trigger modes are modeled.
The FD trigger curves are in good agreement with the MC predictions, the average difference   being -0.5\% for B and -0.7\% for C.\\

{\bf Charge identification.} 
%The identification of the charge $Z$ of a CR particle is based on independent samplings of its specific ionization $dE/dx$ measured with CHD and IMC.
The identification of the particle charge $Z$ is based on the measurements of the ionization deposits in the CHD and IMC. 
The particle trajectory makes it possible  to locate the CHD paddles and IMC fibers traversed by the primary particle and to determine the path length correction 
 to the signals for the extraction of the $dE/dx$ samples. 
Three independent $dE/dx$ measurements are obtained, one for each CHD layer and the third one by averaging the samples (at most eight) along the track in the upper half of the IMC, 
summing up the signals of the crossed fiber in each layer and its two neighbors.
In order to suppress the contribution of possible signals of secondary tracks wrongly associated to the track of the primary nucleus, 
only $dE/dx$ signals larger than 1.5 MeV/mm (corresponding to the energy released in a fiber by 10 MIPs (Minimum Ionizing Particles)) are used in the mean calculation.\\
Calibration curves of $dE/dx$ are built by fitting FD subsets for each nuclear species to a function of $Z^2$ by using a  \enquote{halo} model \cite{GSI},
in which $dE/dx$ is parametrized  as the sum of two contributions (\enquote{core} and  \enquote{halo}, respectively)
\begin{equation}
\frac{dE}{dx} = \frac{A (1 - f_h) \alpha Z^2}{ 1 + B_S (1 - f_h) \alpha Z^2 + C_S \alpha^2 Z^4} + A f_h \alpha Z^2
\label{eq:Voltz}
\end{equation}
where the parameter $f_h$ represents the fraction of energy deposited in the halo;
$B_S$ and $C_S$ model the strength %of the saturation;
 of the scintillation quenching; 
$A$ is an overall normalization constant; and $\alpha$ is close to 2 MeV g$^{-1}$ cm$^2$ for a plastic scintillator.
The parameters are extracted from the fits separately for the CHDX, CHDY and IMC. 
These three calibration curves are then used to reconstruct three charge values ($Z_{\rm{CHDX}}$, $Z_{\rm{CHDY}}$, $Z_{\rm{IMC}}$)
from the measured $dE/dx$ yields on an event-by-event basis.
For high-energy showers, the charge peaks are corrected for a systematic shift 
 to higher values (up to 0.15 $e$) with respect to the nominal charge positions, due to the large amount of shower particle tracks backscattered from the TASC whose signals 
 add up to the primary particle ionization signal. 
The resulting distribution of the reconstructed charge ($Z_{\rm{CHD}}$) combining $Z_{\rm{CHDX}}$ and $Z_{\rm{CHDY}}$ is shown in Fig.~\ref{fig:ZIMCCHD}(a).
B and C candidates are selected by applying  a window cut of half-width 0.45 centered on the nominal charge values  ($Z=5, 6$).
Events with C nuclei undergoing a charge-changing nuclear interaction at the top of the IMC are clearly visible in the tail of the C
drop-shaped distribution extending to lower $Z_{IMC}$ values in Fig.~\ref{fig:ZIMCCHD}(b). 
They are removed  by requiring
consistency between the CHD and IMC charges ($|Z_{\rm{CHD}} - Z_{\rm{IMC}}|< 1$), 
and among the individual charge values measured  in the four upper pairs of adjacent fiber layers.\\ 

{\bf Track width.}
A clustering algorithm is applied to  the fibers being hit in the IMC before track finding and fitting. 
In each IMC layer, neighboring fibers with an energy deposit $>$0.3 MIPs
are clustered around the fibers with larger signals.
The  position of each cluster is computed as the center-of-gravity (COG) of its fibers. 
The cluster positions are taken as candidate track points for the combinatorial Kalman filter algorithm \cite{paolo2017} which is used to 
identify the clusters associated to the primary particle track and to reconstruct its direction and entrance point at the top of the instrument. 
In each layer $l$, we define the  track width as 
\begin{equation}
\rm{TW}_l = \frac{ \sum\limits_{j=m-3}^{m+3} E_{l,j} -  \sum\limits_{j=m-1}^{m+1} E_{l,j} }{Z_l^2}
\end{equation}
where $E_{l,j}$ is the energy deposit in the fiber $j$ of the layer $l$, 
$m$ is the index of the fiber with the maximum signal in the cluster crossed by the primary particle track, 
and the numerator represents the difference between the total energy deposited  in the 7 central fibers of the cluster and 
the cluster core, made of 3 fibers. 
$Z_l$ is the charge in the layer $l$ which is calculated by using the signals of the 5 central fibers in the cluster crossed by the track. 
The total track width is then defined as
\begin{equation}
\rm{TW} =  \frac{1}{6} \sum\limits_{l=1}^{6} \rm{TW}_{l}
\end{equation}
where the sum is limited to the first eight IMC layers from the top excluding the two layers with maximum and minimum $Z_l$, respectively.

The $\rm{TW}$ of interacting events at the top of the instrument is wider than that of penetrating nuclei, 
due to the angular spread of secondary particles produced in the interaction and their lower specific ionization compared to that of the primary particle.
In Fig.~\ref{fig:TW},  a sample of  B events selected in FD by means of the CHD only (i.e. without the IMC consistency cuts described above) 
is compared with the distributions obtained from MC simulations of B
and other nuclei, applying the same selections as for FD. 
It can be noticed that B nuclei traversing CHD and the top of the IMC without interacting show a peak at low $\rm{TW}$ values (red filled histogram), 
while the broad distribution at large $\rm{TW}$ values is  due to the interaction of background particles (mainly protons, He, C) 
misidentified as boron (blue filled histogram). 
A cut on $\rm{TW}<0.18$ is applied to select penetrating B events and reject both early interacting B nuclei (the right-hand tail in the red filled histogram)
and the background from other nuclear species. 

Studying  $\rm{TW}$ distributions similar to the ones shown in Fig.~\ref{fig:TW} but obtained by applying also the IMC consistency cuts, 
a residual  background contamination can be computed as the fraction of  nuclei misidentified as B and not rejected by the $\rm{TW}$ cut, compared
to the number of selected B events in different intervals of $E_{\rm{TASC}}$.
This contamination fraction  is $\sim 1\%$ for $E_{\rm TASC}<10^2$ GeV and grows logarithmically  for $E_{\rm TASC}>10^2$ GeV, 
approaching $\sim 7\%$ at  1.5 TeV. 
The estimated $E_{\rm{TASC}}$ distribution of the background in the final B sample is shown in Fig.~\ref{fig:BC_ETASC} as a blue-filled histogram. 
It is subtracted from the B distribution in FD (red-filled histogram) before the application of the unfolding procedure.\\

{\bf Selection efficiency.}
The efficiency of the complete selection procedure of B and C nuclei, estimated from MC and including trigger, tracking, charge identification and $\rm{TW}$ efficiencies, 
is shown as a function of the kinetic energy per nucleon in  Fig.~\ref{fig:BCeff}.\\

\textcolor{black}{{\bf Systematic uncertainties.}
The flux systematic relative errors stemming from several sources, 
including HE trigger efficiency, charge identification, MC simulations, energy scale, energy unfolding,  background contamination,  and live time
are shown in Fig.~\ref{fig:sys_all} as a function of the kinetic energy per nucleon.
The dominant source of uncertainty in the flux derives from  the different predictions of  the energy response matrix  by simulations based on EPICS and Geant4.
Energy-independent systematic uncertainties affecting the flux normalization include live time (3.4\%), long-term stability of charge calibration (0.5\%), 
energy scale calibration (3\%), and assumption of the B isotopic composition (1.7\%).
With the exception of the latter, the other energy-independent systematic uncertainties cancel out completely in the B/C ratio. }

\textcolor{black}{
Additional systematic effects that have been studied extensively are related to the particle interactions in the materials of the instrument. 
Primary particles cross a  2 mm-think Al panel covering the top of the instrument, before reaching the CHD. The probability of interactions of B and C nuclei in this panel is $\lesssim$1\%.
This effect is taken into account in the flux calculation.\\}
\textcolor{black}{CR nuclei traverse several materials in the IMC, mainly composed of CFRP (Carbon Fiber Reinforced Polymer), aluminum and tungsten. 
Possible uncertainties in the inelastic cross sections in MC simulations or discrepancies in the material description might affect the flux normalization. 
We have checked that hadronic interactions are well simulated in the detector, by measuring the survival probabilities of C nuclei at different depths in the IMC.
The survival probabilities are in agreement with MC prediction within $<$1\% as shown in Fig.~\ref{fig:Csurv}.
}

\textcolor{black}{Several studies were performed to check the stability of the detector performance.  
Day-by-day calibrations of the detector channels are performed by using penetrating protons and He particles selected by a dedicated trigger mode.
This ensures that the CHD and IMC charge measurements are stable over time at the level of 0.5\%.\\
To investigate the uncertainty in the definition of the acceptance, restricted acceptance (up to 20\% of the nominal one) regions were also studied. The corresponding fluxes are consistent within statistical fluctuations.\\
To investigate possible  time-dependent effects in the energy scale of the TASC, we have compared C flux measurements obtained with subsets of data taken in different periods of time.
We have chosen for comparison an energy interval between 30 GeV/$n$ and 300 GeV/$n$, to exclude the low-energy region where the flux is affected by solar modulation and 
the high-energy region where statistical fluctuations are relevant. 
In this energy interval,  the fluxes in different time periods  turned out to be in agreement at a level consistent with the energy scale calibration error (3\%). \\}

\textcolor{black}{{\bf Energy spectra.}
The CALET energy spectra of B and C and the B/C flux ratio, together with a compilation of the available data, are shown in Fig.~\ref{fig:flux}, 
which is an enlarged version of Fig.~2 in the main body of the paper.
In tables  \ref{tab:Bflux} and \ref{tab:Cflux}, 
the B and C differential fluxes  in different energy intervals are reported with the separate contributions to the flux error of the statistical uncertainties, 
the systematic uncertainties in normalization, and energy dependent systematic uncertainties.
The data of the B/C flux ratio are reported in  table  \ref{tab:BCratio}.
}
%%%%%%%%%%%%%%%%%%%%%%%%%%%%%%%%%%%%%%%%%%%%%%%%%%%%
\clearpage
\begin{figure}[hbt!] \centering
\subfigure[]
{
\includegraphics[scale=0.65]{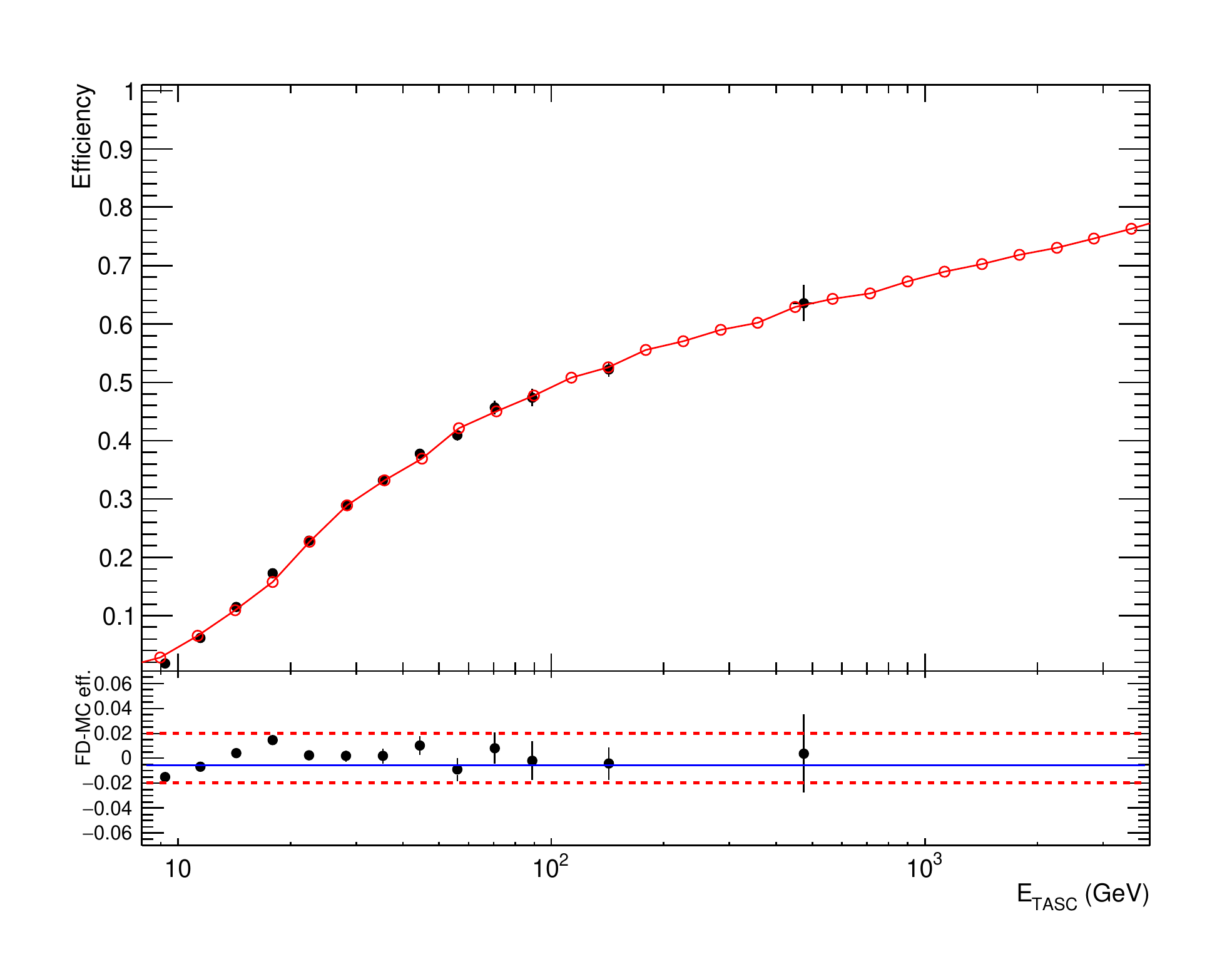}
\label{fig:HEeffC}                            
}
\subfigure[]
{
\includegraphics[scale=0.65]{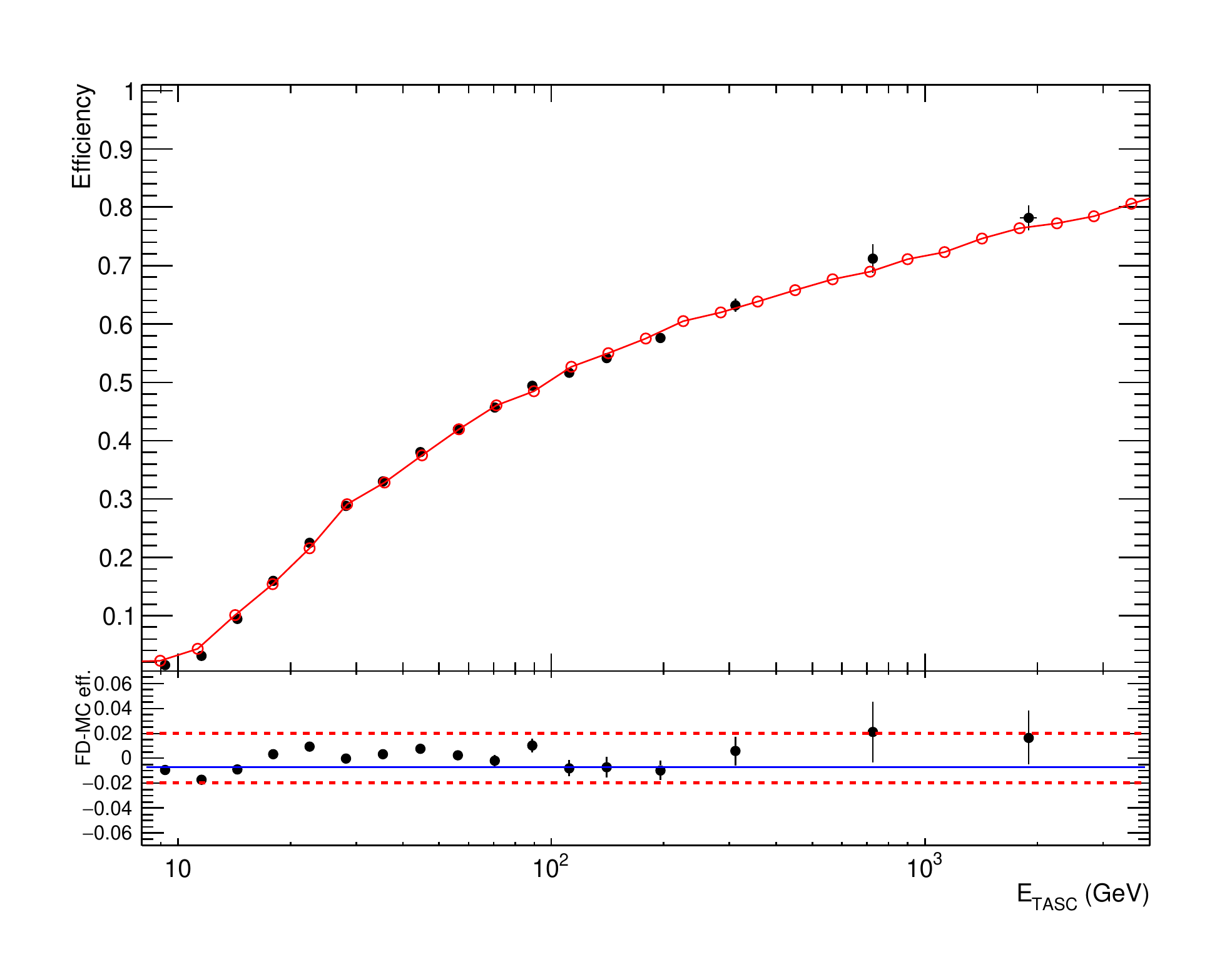}
\label{fig:HEeffB}                            
}
\caption{HE trigger efficiency as a function of the deposited energy in the TASC for B ({\it a}) and C ({\it b}) as derived from flight data (FD) (black dots) and Monte Carlo (MC) (red curves). 
The difference between FD and MC efficiencies  is within $\pm 2\%$ over the whole energy range, as shown by the red dotted lines in the bottom plots of  panels ({\it a}) and ({\it b}).
The average difference is -0.5\% for B and -0.7\% for C, as indicated by the blue lines. } 
\label{fig:HEeff}
\end{figure}\noindent%
\clearpage
%%%%%%%%%%%%%%%%%%%%%%%%%%%
\begin{figure}
\begin{center}
%\vspace{2mm}
\subfigure[]
{
\includegraphics[scale=0.6]{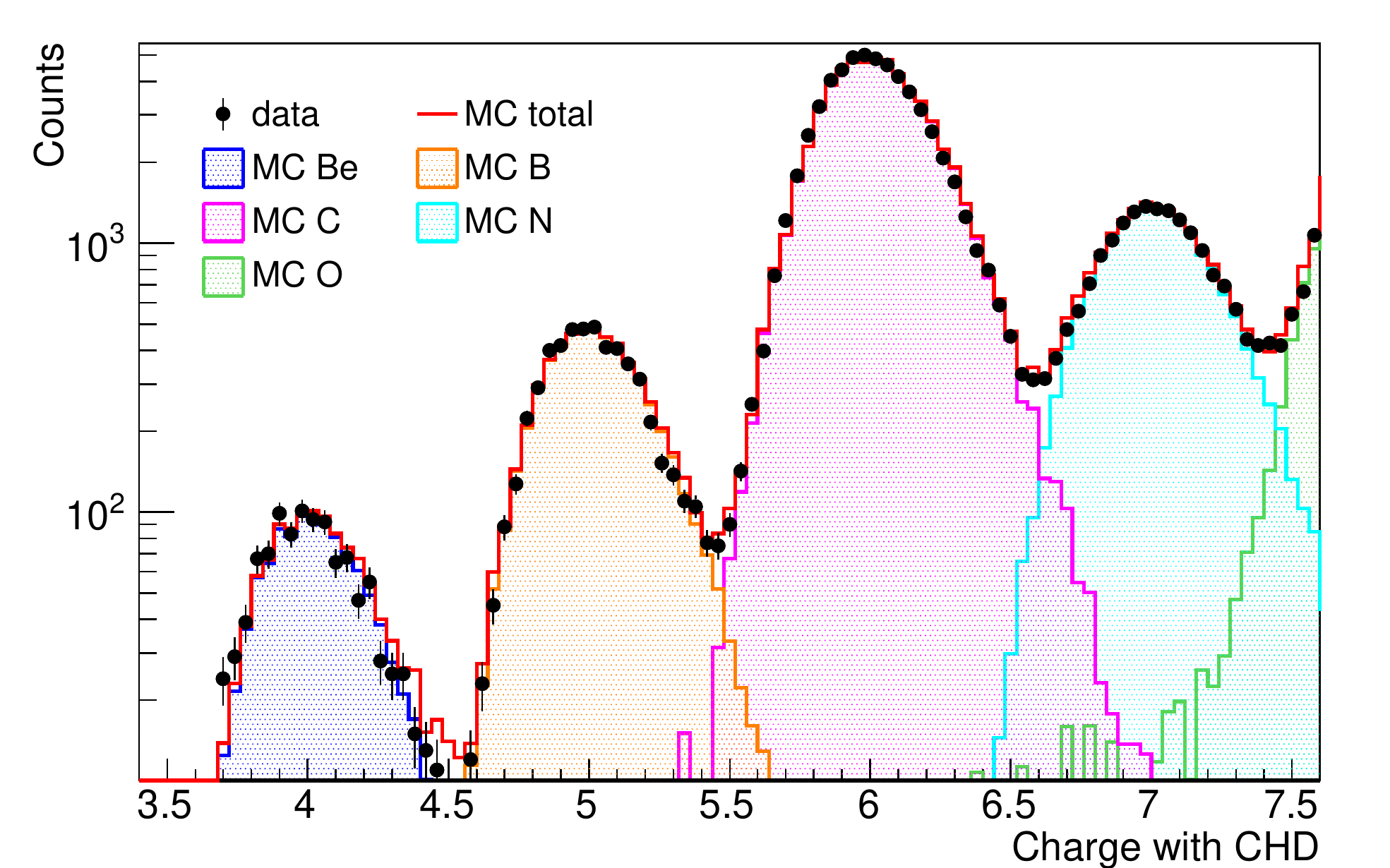}
\label{fig:charge1}                            
}
\subfigure[]
{
\includegraphics[scale=0.47]{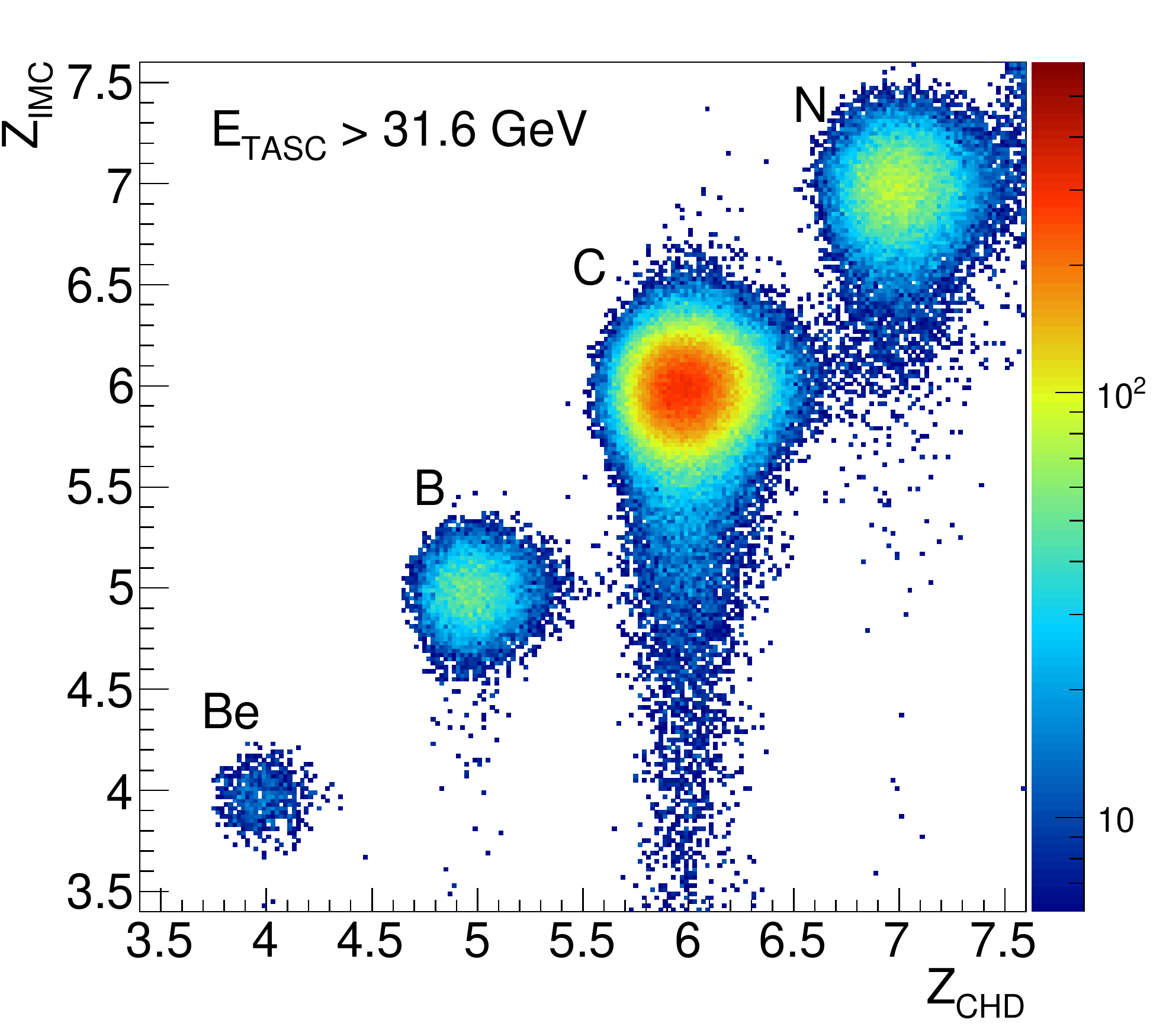}
\label{fig:charge2}                            
}
%\begin{figure} \centering
%\includegraphics[scale=0.7]{BCNOcharge.eps}
\caption{
%Charge distribution from the combined CHD layers 
 ({\it a})  Charge distribution measured in  the energy interval 100 $< E_{\rm{TASC}}/\text{GeV} < $ 215 by the combined CHD layers  (FD, black dots) compared to MC simulations.
Events are selected  by requiring a measured charge in IMC consistent with $Z_{\rm CHD}$ and a track width $\rm{TW}<0.18$ (Fig.~\ref{fig:TW}).
({\it b})  Charge correlation between  $Z_{\rm{IMC}}$ and $Z_{\rm{CHD}}$  in a sample of FD selected
without applying the consistency cuts between CHD and IMC or the $\rm{TW}$ cut.}%in the elemental range between B and F.  
\label{fig:ZIMCCHD}
\end{center}
\end{figure}\noindent
%%%%%%%%%%%%%%%%%%%%%%%%%%%%%%%%%%%%%%%%%%%%%%%%%%%%%%%%%%%%%%%%%%%%%%%%%%%
\begin{figure}
\begin{center}
%\vspace{2mm}
\subfigure[]
{
\includegraphics[scale=0.7]{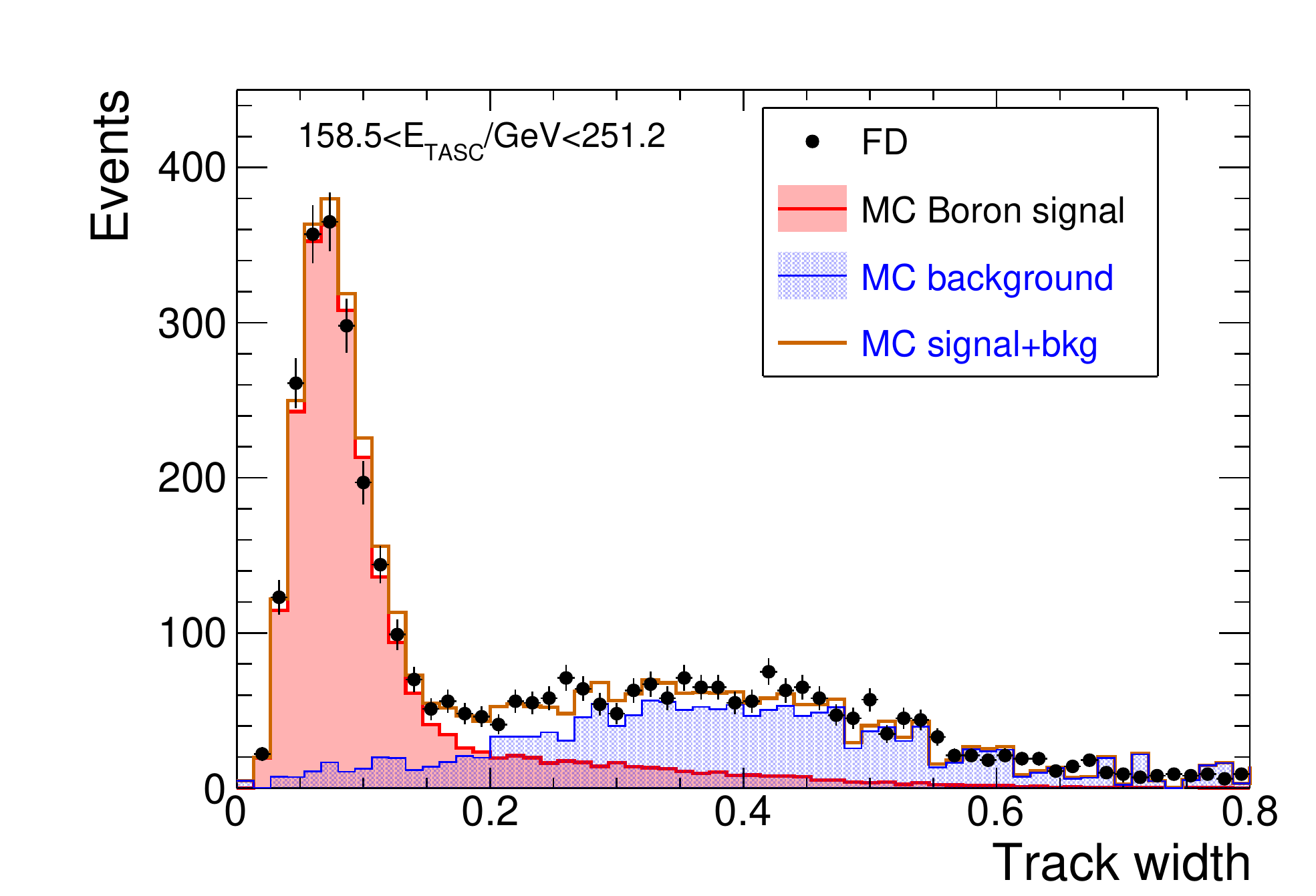}
\label{fig:TW1}                            
}
\subfigure[]
{
\includegraphics[scale=0.7]{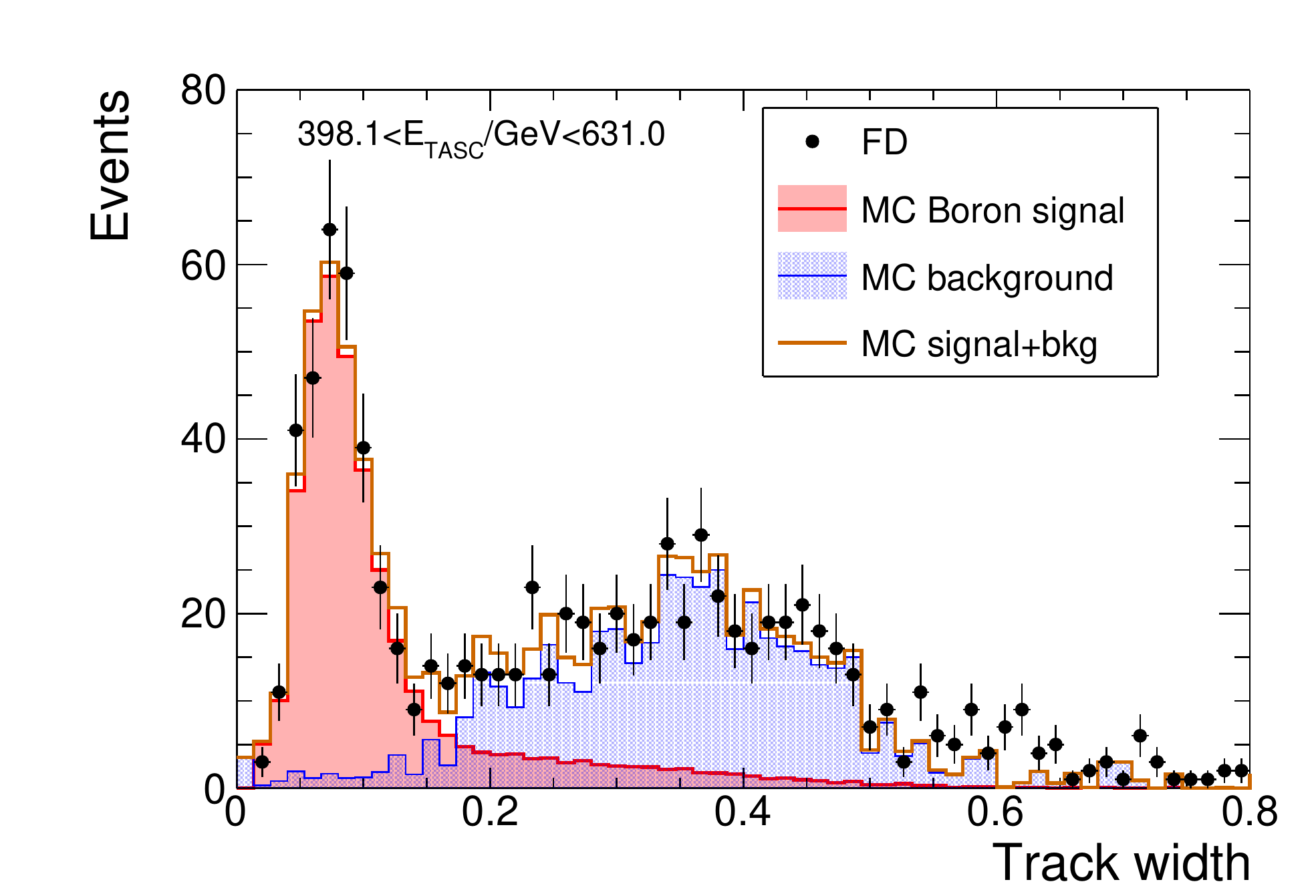}
\label{fig:TW2}                            
}
\caption{Distributions of the track width ($\rm{TW}$) variable in two different intervals of  $E_{\rm TASC}$. 
The black dots represent a sample of  B events selected in FD by means of the CHD only. 
MC distributions of B and other background nuclei (mainly proton, He, C) are obtained with the same selections used for FD. }
\label{fig:TW}
\end{center}
\end{figure}
%%%%%%%%%%%%%%%%%%%%%%%%%%%%%%%%%%%%%%%%%%%%%%%%%%%%%%%%%%%%%%
\clearpage
\begin{figure}
\begin{center}
\includegraphics[scale=0.6]{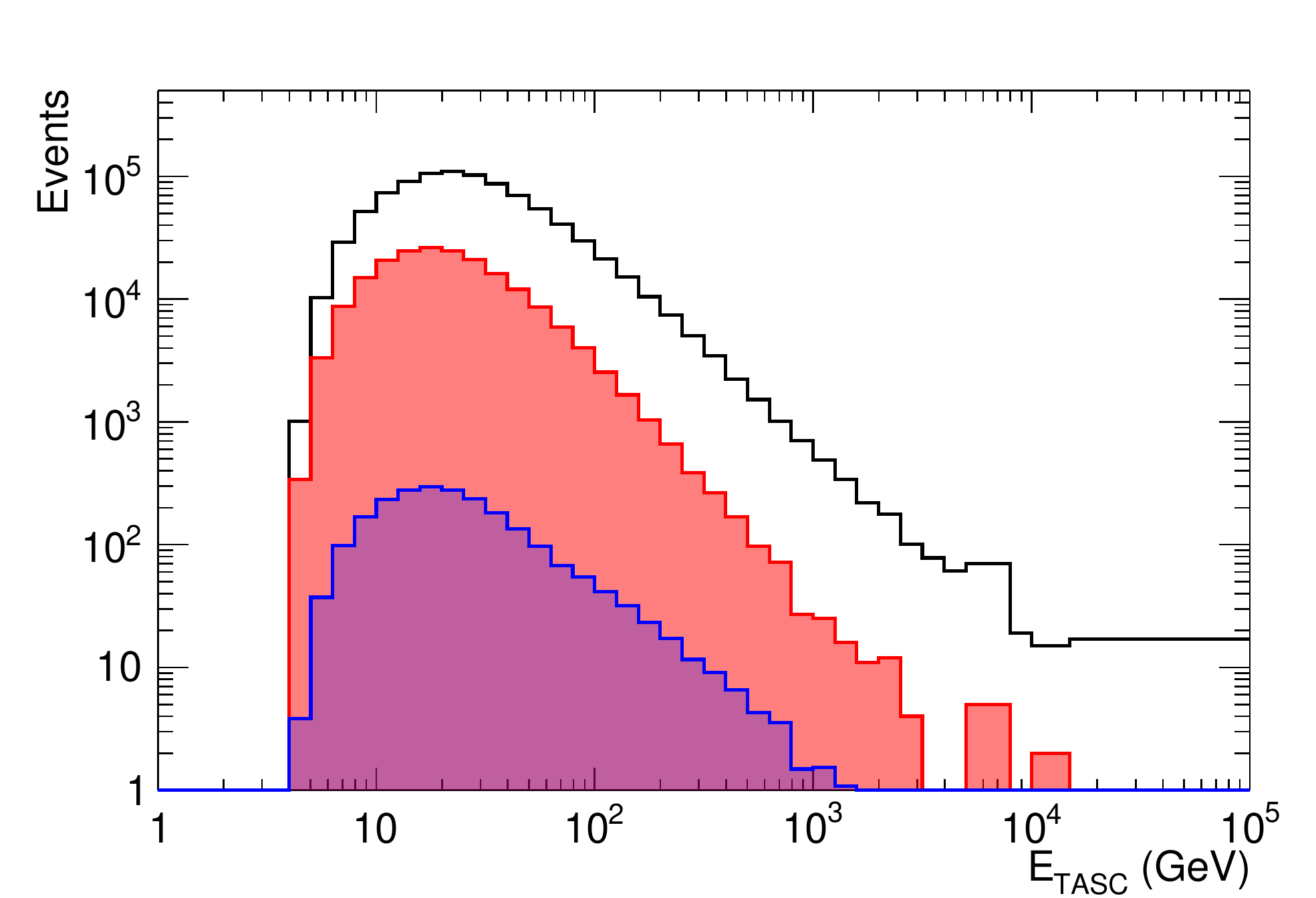}
\caption{Distribution of total deposited energy in the TASC for the final selection of carbon (black histogram) and boron (red histogram) candidates in FD. 
The blue-filled histogram represents the background contamination  subtracted from the B sample for the flux measurement.}
%The background contamination  is estimated from the fraction of events with $TW>0.18$ in $TW$ distributions obtained after applying the complete charge selection procedure for B to FD.
\label{fig:BC_ETASC}
\end{center}
\end{figure}\noindent
%%%%%%%%%%%%%%%%%%%%%%%%%%%%%%%%%%%%%%%%%%%%%%%%%%%%%%%%%%%%%%
\begin{figure}
%\vspace{0.4cm}
\begin{center}
\includegraphics[scale=0.58]{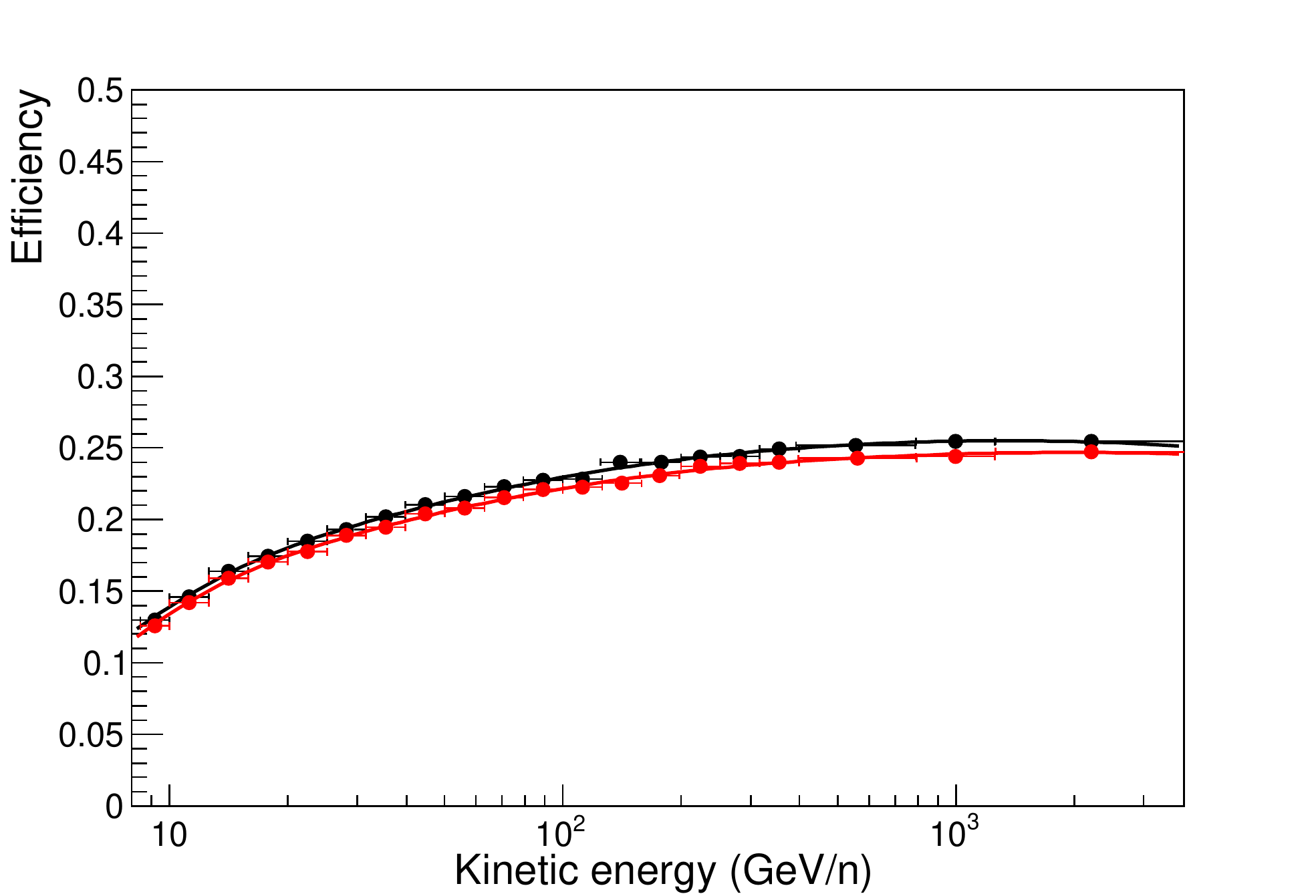}
\caption{Total selection efficiency for B (red dots) and C (black dots) estimated with MC simulations.}
\label{fig:BCeff}
\end{center}
\end{figure}\noindent
%%%%%%%%%%%%%%%%%%%%%%%%%%%%%%%%%%%%%%%%%%%%%%%%%%%%%%%%%%%%%%%
\clearpage
%%%%%%%%%%%%%%%%%%%%%%%%%%%%%%%%%%%%%%%%
%%%%%%%%%%%%%%%%%%%%%%%%%%%%%%%%%%%%%%%%
\clearpage
\begin{figure}[hbt!] \centering
\subfigure[]
{
  \includegraphics[width=15cm, height=9cm]{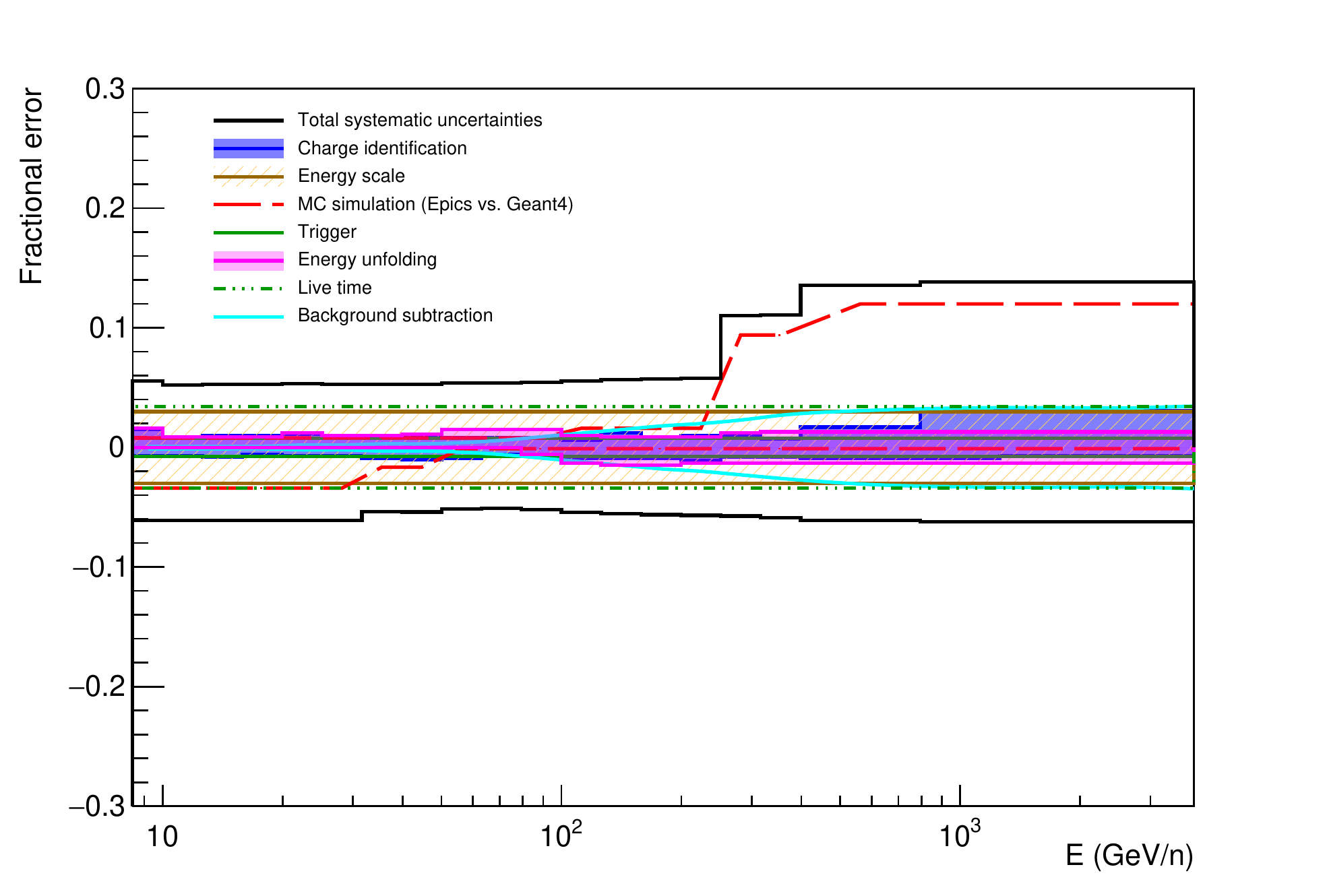}
  \label{fig:sysB}                            
}
\subfigure[]
{
  \includegraphics[width=15cm, height=9cm]{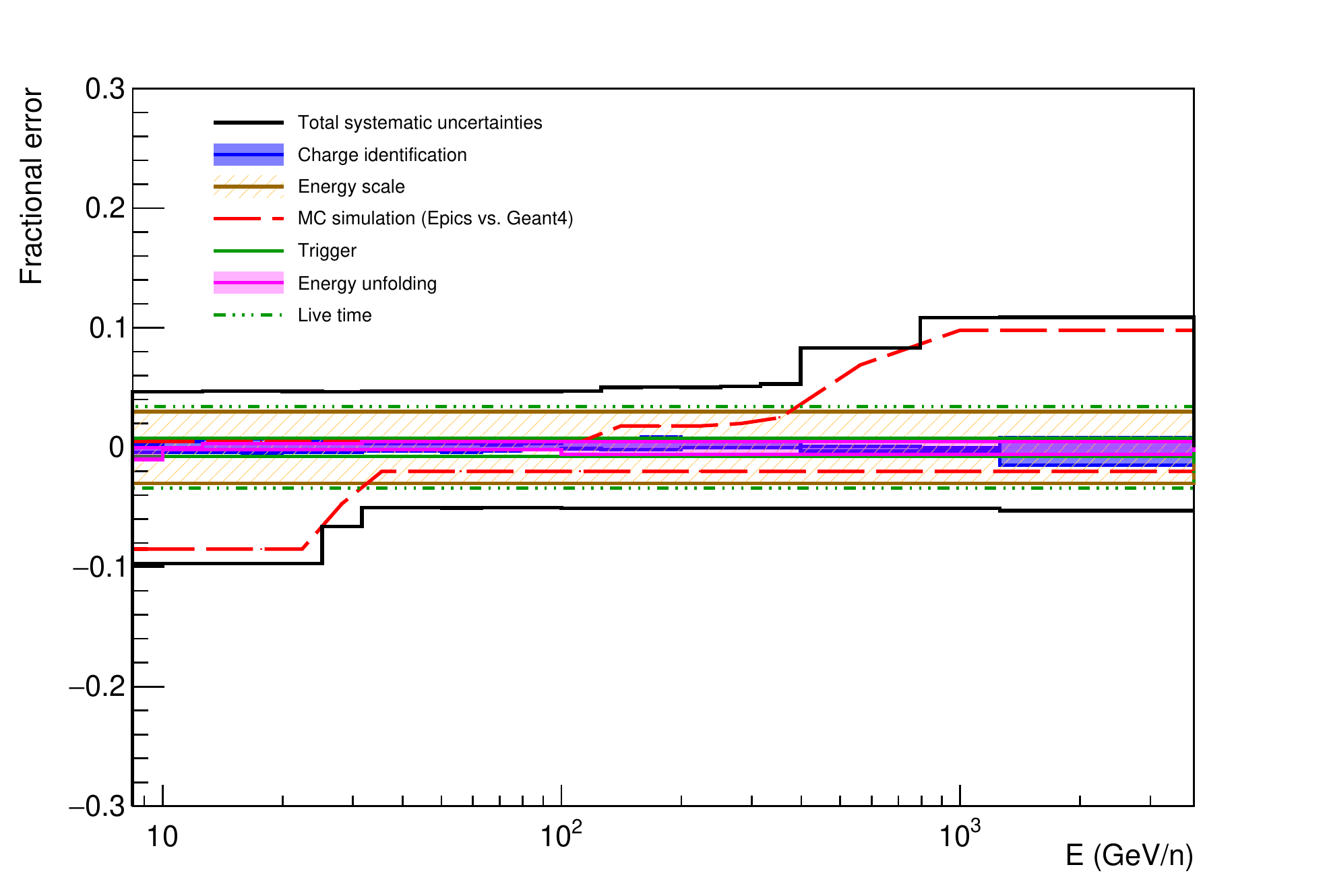}
  \label{fig:sysC}                            
}
\caption{Energy dependence of  systematic uncertainties (relative errors) for B  ({\it a}) and C ({\it b}).  
 The band within the black lines shows the sum in quadrature of all the sources of systematics. 
 A detailed breakdown of systematic errors  stemming from charge identification, HE trigger, MC simulations, energy scale, energy unfolding, background contamination, live time  is shown.
}
\label{fig:sys_all}
\end{figure}\noindent
%%%%%%%%%%%%%%%%%%%%%%%%%%%%%%%%%%%%%%%%%%%%%%%%%%%%%%%%%%%%%%
\begin{figure}
\begin{center}
\includegraphics[scale=0.58]{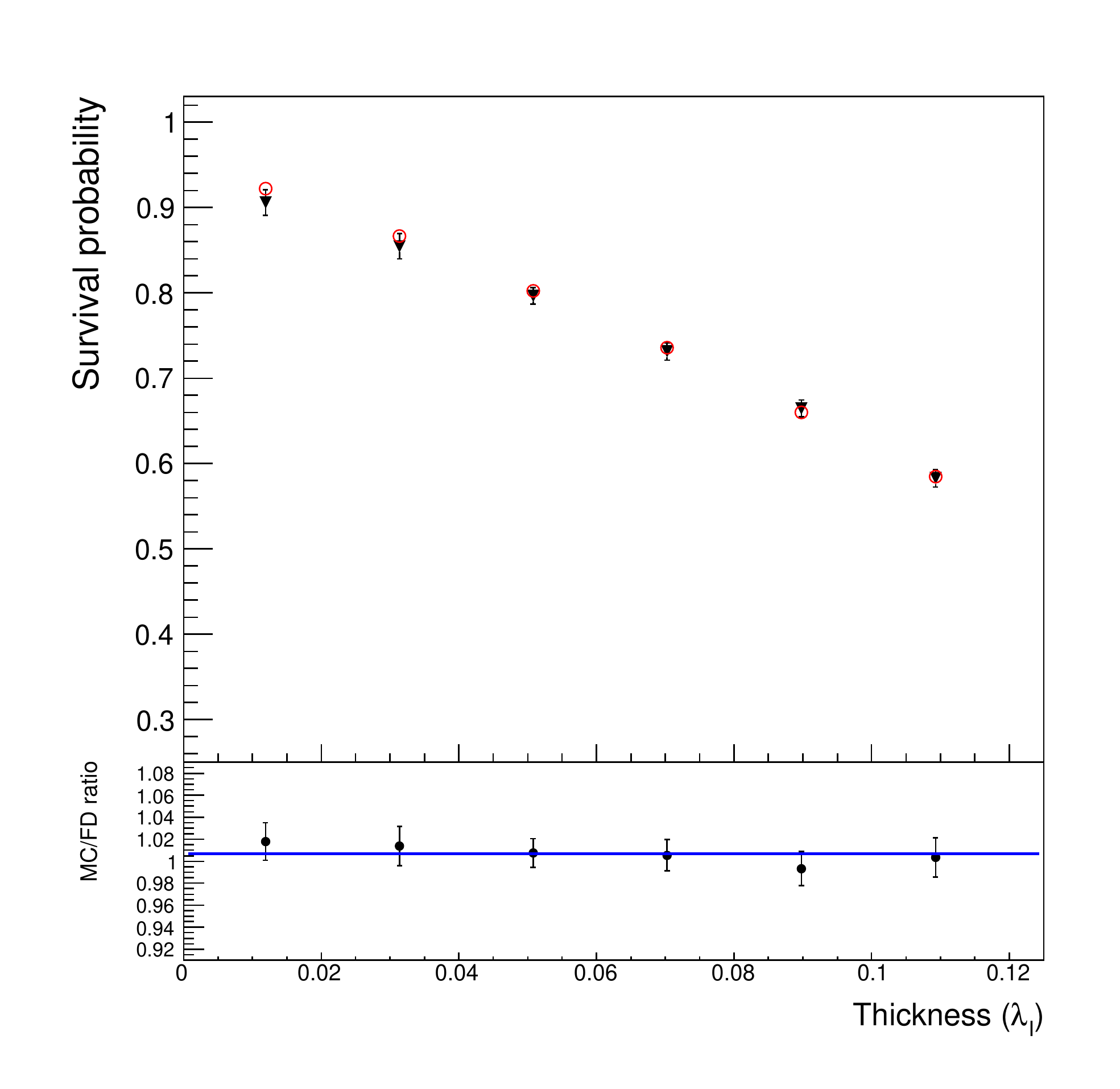}
\caption{Survival probability  as a function of  the material thickness traversed by C nuclei  in the IMC
as derived from FD (black triangles) and MC (red circles).  The survival probabilities are calculated in FD  by dividing the number of events 
selected as C  in the first six double layers of scintillating fibers in the IMC by the number of C  events selected with the CHD. 
The material thickness is expressed in units of proton interaction length $\lambda_I$ and it is measured from the bottom of the CHD. 
The double fiber layers are preceded by 0.2 $X_0$-thick W plates (except the first pair), Al honeycomb panels and CFRP supporting structures. 
%The red line is a fit to the data with an exponential function. 
In the bottom panel, the blue  line represents a constant value fitted to the ratio between MC and FD survival probabilities. 
The fitted value is $1.006\pm0.006$.}
\label{fig:Csurv}
\end{center}
\end{figure}\noindent
%%%%%%%%%%%%%%%%%%%%%%%%%%%%%%%%%%%%%%%%%%%%%%%%%%%%%%%%%%%%%%%
%%%%%%%%%%%%%%%%%%%%%%%%%%%%%%%%%%%%%%%%%%%%
\clearpage
\begin{figure}[hbt!] \centering
\includegraphics[width=0.75\hsize]{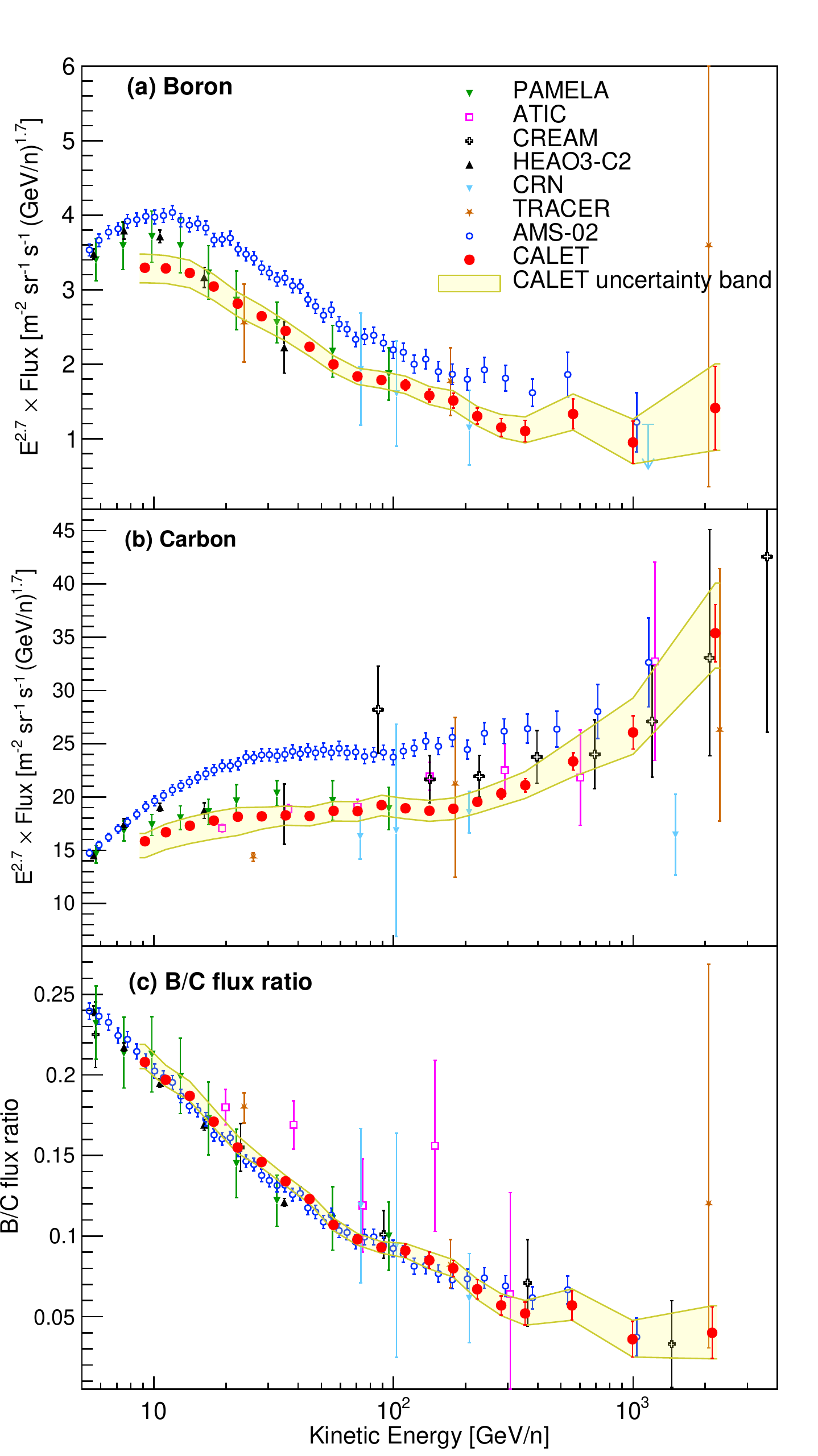} 
\caption{CALET (a) boron and (b) carbon flux (multiplied by $E^{2.7}$) and (c)  ratio of boron to carbon, as a function of kinetic energy $E$.
Error bars of CALET data (red) represent the statistical uncertainty only, while the yellow band indicates the quadratic sum of statistical and systematic errors. 
Also plotted are other direct measurements \cite{HEAO, CRN-BC, CREAM1-BC, TRACER2012, PAMELA-C, AMS-PhysRep, AMS-LiBeB, ATIC, CREAM2}.}
\label{fig:flux}
\end{figure}
\noindent
%%%%%%%%%%%%%%%%%%%%%%%%%
\clearpage
\renewcommand{\arraystretch}{1.25}
\begin{table*}
  \caption{Table of CALET  boron spectrum. 
 The first, second and third errors in the flux represent the statistical uncertainties, systematic uncertainties in normalization, and energy dependent systematic uncertainties, respectively.
\label{tab:Bflux}}
\begin{ruledtabular}
\begin{tabular}{c c c c}
Energy Bin [GeV$/n$] & Flux [m$^{-2}$sr$^{-1}$s$^{-1}$(GeV$/n$)$^{-1}$]   \\
%(GeV/n)  & (m$^{-2}$sr$^{-1}$s$^{-1}$GeV$^{-1}$) &  \\
\hline
   8.4--10.0 & $(8.313 \, \pm 0.045 \,  \pm 0.405 \, _{-0.307}^{+0.221}) \times 10^{-3}$ \\ 
  10.0--12.6 & $(4.811 \, \pm 0.029 \,  \pm 0.234 \, _{-0.178}^{+0.092}) \times 10^{-3}$ \\ 
  12.6--15.8 & $(2.535 \, \pm 0.018 \,  \pm 0.123 \, _{-0.094}^{+0.051}) \times 10^{-3}$ \\ 
  15.8--20.0 & $(1.285 \, \pm 0.011 \,  \pm 0.063 \, _{-0.047}^{+0.026}) \times 10^{-3}$ \\ 
  20.0--25.1 & $(6.379 \, \pm 0.065 \,  \pm 0.311 \, _{-0.236}^{+0.142}) \times 10^{-4}$ \\ 
  25.1--31.6 & $(3.219 \, \pm 0.039 \,  \pm 0.157 \, _{-0.119}^{+0.065}) \times 10^{-4}$ \\ 
  31.6--39.8 & $(1.600 \, \pm 0.024 \,  \pm 0.078 \, _{-0.037}^{+0.033}) \times 10^{-4}$ \\ 
  39.8--50.1 & $(7.846 \, \pm 0.142 \,  \pm 0.382 \, _{-0.182}^{+0.164}) \times 10^{-5}$ \\ 
  50.1--63.1 & $(3.768 \, \pm 0.085 \,  \pm 0.183 \, _{-0.060}^{+0.087}) \times 10^{-5}$ \\ 
  63.1--79.4 & $(1.859 \, \pm 0.052 \,  \pm 0.091 \, _{-0.028}^{+0.044}) \times 10^{-5}$ \\ 
  79.4--100.0 & $(9.723 \, \pm 0.335 \,  \pm 0.473 \, _{-0.176}^{+0.237}) \times 10^{-6}$ \\ 
 100.0--125.9 & $(5.026 \, \pm 0.210 \,  \pm 0.245 \, _{-0.118}^{+0.133}) \times 10^{-6}$ \\ 
 125.9--158.5 & $(2.476 \, \pm 0.130 \,  \pm 0.121 \, _{-0.065}^{+0.073}) \times 10^{-6}$ \\ 
 158.5--199.5 & $(1.273 \, \pm 0.083 \,  \pm 0.062 \, _{-0.036}^{+0.038}) \times 10^{-6}$ \\ 
 199.5--251.2 & $(5.889 \, \pm 0.483 \,  \pm 0.287 \, _{-0.172}^{+0.186}) \times 10^{-7}$ \\ 
 251.2--316.2 & $(2.795 \, \pm 0.290 \,  \pm 0.136 \, _{-0.085}^{+0.276}) \times 10^{-7}$ \\ 
 316.2--398.1 & $(1.438 \, \pm 0.189 \,  \pm 0.070 \, _{-0.047}^{+0.143}) \times 10^{-7}$ \\ 
 398.1--794.3 & $(5.010 \, \pm 0.753 \,  \pm 0.244 \, _{-0.186}^{+0.633}) \times 10^{-8}$ \\ 
 794.3--1258.9 & $(7.553 \, \pm 2.253 \,  \pm 0.368 \, _{-0.292}^{+0.978}) \times 10^{-9}$ \\ 
1258.9--3860.5 & $(1.328 \, \pm 0.528 \,  \pm 0.065 \, _{-0.051}^{+0.172}) \times 10^{-9}$ \\ 
\end{tabular}
\end{ruledtabular}
\end{table*}
\renewcommand{\arraystretch}{1.0}

%\clearpage
\renewcommand{\arraystretch}{1.25}
\begin{table*}
  \caption{Table of CALET carbon spectrum. 
  The first, second and third errors in the flux represent the statistical uncertainties, systematic uncertainties in normalization, and energy dependent systematic uncertainties, respectively.
\label{tab:Cflux}}
\begin{ruledtabular}
\begin{tabular}{c c c}
Energy Bin [GeV$/n$] & Flux [m$^{-2}$sr$^{-1}$s$^{-1}$(GeV$/n$)$^{-1}$]   \\
%(GeV/n)  & (m$^{-2}$sr$^{-1}$s$^{-1}$GeV$^{-1}$) &  \\
\hline
   8.4--10.0 & $(3.990 \, \pm 0.010 \,  \pm 0.182 \, _{-0.343}^{+0.037}) \times 10^{-2}$ \\ 
  10.0--12.6 & $(2.440 \, \pm 0.006 \,  \pm 0.111 \, _{-0.208}^{+0.025}) \times 10^{-2}$ \\ 
  12.6--15.8 & $(1.358 \, \pm 0.004 \,  \pm 0.062 \, _{-0.116}^{+0.016}) \times 10^{-2}$ \\ 
  15.8--20.0 & $(7.496 \, \pm 0.026 \,  \pm 0.342 \, _{-0.641}^{+0.088}) \times 10^{-3}$ \\ 
  20.0--25.1 & $(4.108 \, \pm 0.016 \,  \pm 0.187 \, _{-0.351}^{+0.044}) \times 10^{-3}$ \\ 
  25.1--31.6 & $(2.211 \, \pm 0.010 \,  \pm 0.101 \, _{-0.106}^{+0.023}) \times 10^{-3}$ \\ 
  31.6--39.8 & $(1.193 \, \pm 0.006 \,  \pm 0.054 \, _{-0.026}^{+0.013}) \times 10^{-3}$ \\ 
  39.8--50.1 & $(6.385 \, \pm 0.040 \,  \pm 0.291 \, _{-0.138}^{+0.069}) \times 10^{-4}$ \\ 
  50.1--63.1 & $(3.520 \, \pm 0.025 \,  \pm 0.161 \, _{-0.077}^{+0.038}) \times 10^{-4}$ \\ 
  63.1--79.4 & $(1.888 \, \pm 0.016 \,  \pm 0.086 \, _{-0.041}^{+0.020}) \times 10^{-4}$ \\ 
  79.4--100.0 & $(1.045 \, \pm 0.011 \,  \pm 0.048 \, _{-0.023}^{+0.011}) \times 10^{-4}$ \\ 
 100.0--125.9 & $(5.523 \, \pm 0.067 \,  \pm 0.252 \, _{-0.123}^{+0.063}) \times 10^{-5}$ \\ 
 125.9--158.5 & $(2.930 \, \pm 0.043 \,  \pm 0.134 \, _{-0.065}^{+0.062}) \times 10^{-5}$ \\ 
 158.5--199.5 & $(1.589 \, \pm 0.028 \,  \pm 0.073 \, _{-0.035}^{+0.035}) \times 10^{-5}$ \\ 
 199.5--251.2 & $(8.830 \, \pm 0.186 \,  \pm 0.403 \, _{-0.196}^{+0.185}) \times 10^{-6}$ \\ 
 251.2--316.2 & $(4.931 \, \pm 0.122 \,  \pm 0.225 \, _{-0.109}^{+0.112}) \times 10^{-6}$ \\ 
 316.2--398.1 & $(2.750 \, \pm 0.082 \,  \pm 0.125 \, _{-0.061}^{+0.074}) \times 10^{-6}$ \\ 
 398.1--794.3 & $(8.770 \, \pm 0.305 \,  \pm 0.400 \, _{-0.196}^{+0.610}) \times 10^{-7}$ \\ 
 794.3--1258.9 & $(2.070 \, \pm 0.124 \,  \pm 0.094 \, _{-0.046}^{+0.204}) \times 10^{-7}$ \\ 
1258.9--3860.5 & $(3.325 \, \pm 0.252 \,  \pm 0.152 \, _{-0.089}^{+0.328}) \times 10^{-8}$ \\
 \end{tabular}
\end{ruledtabular}
\end{table*}
\renewcommand{\arraystretch}{1.0}

\clearpage
\renewcommand{\arraystretch}{1.25}
\begin{table*}
  \caption{Table of CALET boron to carbon flux ratio. 
  The first and second  errors represent the statistical uncertainties and systematic uncertainties, respectively.
\label{tab:BCratio}}
\begin{ruledtabular}
\begin{tabular}{c c c}
Energy Bin [GeV$/n$] & B/C    \\
\hline
%   8.4--10.0 & $ 0.208 \, \pm 0.001 \, _{-0.005}^{+0.010}$  \\ 
%  10.0--12.6 & $ 0.197 \, \pm 0.001 \, _{-0.004}^{+0.009}$  \\ 
 % 12.6--15.8 & $ 0.187 \, \pm 0.001 \, _{-0.004}^{+0.008}$  \\ 
 % 15.8--20.0 & $ 0.171 \, \pm 0.002 \, _{-0.004}^{+0.008}$  \\ 
%  20.0--25.1 & $ 0.155 \, \pm 0.002 \, _{-0.003}^{+0.007}$  \\ 
%  25.1--31.6 & $ 0.146 \, \pm 0.002 \, _{-0.003}^{+0.006}$  \\ 
 % 31.6--39.8 & $ 0.134 \, \pm 0.002 \, _{-0.003}^{+0.006}$  \\ 
%  39.8--50.1 & $ 0.123 \, \pm 0.002 \, _{-0.003}^{+0.005}$  \\ 
%  50.1--63.1 & $ 0.107 \, \pm 0.003 \, _{-0.003}^{+0.005}$  \\ 
%  63.1--79.4 & $ 0.098 \, \pm 0.003 \, _{-0.002}^{+0.004}$  \\ 
%  79.4--100.0 & $ 0.093 \, \pm 0.003 \, _{-0.002}^{+0.004}$  \\ 
% 100.0--125.9 & $ 0.091 \, \pm 0.004 \, _{-0.002}^{+0.005}$  \\ 
% 125.9--158.5 & $ 0.085 \, \pm 0.005 \, _{-0.002}^{+0.005}$  \\ 
% 158.5--199.5 & $ 0.080 \, \pm 0.005 \, _{-0.002}^{+0.004}$  \\ 
% 199.5--251.2 & $ 0.067 \, \pm 0.006 \, _{-0.002}^{+0.004}$  \\ 
% 251.2--316.2 & $ 0.057 \, \pm 0.006 \, _{-0.002}^{+0.003}$  \\ 
% 316.2--398.1 & $ 0.052 \, \pm 0.007 \, _{-0.002}^{+0.007}$  \\ 
% 398.1--794.3 & $ 0.057 \, \pm 0.009 \, _{-0.002}^{+0.008}$  \\ 
% 794.3--1258.9 & $ 0.036 \, \pm 0.011 \, _{-0.001}^{+0.005}$  \\ 
%1258.9--3860.5 & $ 0.040 \, \pm 0.016 \, _{-0.002}^{+0.005}$  \\ 
   8.4--10.0 & $ 0.208 \, \pm 0.001 \, _{-0.004}^{+0.011}$  \\ 
  10.0--12.6 & $ 0.197 \, \pm 0.001 \, _{-0.004}^{+0.009}$  \\ 
  12.6--15.8 & $ 0.187 \, \pm 0.001 \, _{-0.003}^{+0.009}$  \\ 
  15.8--20.0 & $ 0.171 \, \pm 0.002 \, _{-0.003}^{+0.008}$  \\ 
  20.0--25.1 & $ 0.155 \, \pm 0.002 \, _{-0.003}^{+0.007}$  \\ 
  25.1--31.6 & $ 0.146 \, \pm 0.002 \, _{-0.003}^{+0.003}$  \\ 
  31.6--39.8 & $ 0.134 \, \pm 0.002 \, _{-0.002}^{+0.003}$  \\ 
  39.8--50.1 & $ 0.123 \, \pm 0.002 \, _{-0.002}^{+0.003}$  \\ 
  50.1--63.1 & $ 0.107 \, \pm 0.003 \, _{-0.002}^{+0.003}$  \\ 
  63.1--79.4 & $ 0.098 \, \pm 0.003 \, _{-0.002}^{+0.002}$  \\ 
  79.4--100.0 & $ 0.093 \, \pm 0.003 \, _{-0.002}^{+0.002}$  \\ 
 100.0--125.9 & $ 0.091 \, \pm 0.004 \, _{-0.002}^{+0.002}$  \\ 
 125.9--158.5 & $ 0.085 \, \pm 0.005 \, _{-0.002}^{+0.002}$  \\ 
 158.5--199.5 & $ 0.080 \, \pm 0.005 \, _{-0.002}^{+0.002}$  \\ 
 199.5--251.2 & $ 0.067 \, \pm 0.006 \, _{-0.002}^{+0.002}$  \\ 
 251.2--316.2 & $ 0.057 \, \pm 0.006 \, _{-0.002}^{+0.004}$  \\ 
 316.2--398.1 & $ 0.052 \, \pm 0.007 \, _{-0.002}^{+0.004}$  \\ 
 398.1--794.3 & $ 0.057 \, \pm 0.009 \, _{-0.002}^{+0.005}$  \\ 
 794.3--1258.9 & $ 0.036 \, \pm 0.011 \, _{-0.001}^{+0.004}$  \\ 
1258.9--3860.5 & $ 0.040 \, \pm 0.016 \, _{-0.002}^{+0.005}$  \\ 
\hline
\end{tabular}
\end{ruledtabular}
\end{table*}
\renewcommand{\arraystretch}{1.0}

\providecommand{\noopsort}[1]{}\providecommand{\singleletter}[1]{#1}%

\end{document}